\documentclass{mn2e} 
\usepackage{psfig}
\usepackage{times}

\newif\ifAMStwofonts

\title[A synthesis model for AGN evolution]
      {A synthesis model for AGN evolution: supermassive black holes growth and feedback modes}

\author[Merloni and Heinz] {Andrea Merloni$^{1}$ and 
Sebastian Heinz$^{2}$
\\$^{1}$Max-Planck-Institut f\"ur Extraterrestrische Physik,
Giessenbachstr., D-85741, Garching, Germany
\\$^{2}$Astronomy
Department, University of Wisconsin-Madison, Madison, WI 53706}
\date{}

\begin{document}

\maketitle

\label{firstpage}

\begin{abstract}
We present a comprehensive synthesis model for the AGN evolution and the
growth of supermassive black hole in the Universe. We assume that black
holes accrete in just three distinct physical states, or ``modes'': at
low Eddington ratio, only a radiatively inefficient, 
kinetically dominated mode is allowed (LK); at high Eddington
ratio, instead, AGN may display both a purely radiative (radio quiet,
HR), and a kinetic (radio loud, HK) mode. We solve the
continuity equation for the black hole mass function using the locally
determined one as a boundary condition, and the 
hard X-ray luminosity function as tracer of the AGN growth rate
distribution, supplemented with a luminosity-dependent bolometric
correction and an absorbing column distribution. Differently from most
previous semi-analytic and numerical models for black hole growth, we
do not assume any specific distribution of Eddington ratios, rather we
determine it empirically by coupling the mass and luminosity
functions and the set of fundamental relations between observables in
the three accretion modes. SMBH always show a very broad accretion rate
distribution, and we discuss the profound consequences of this fact
for our understanding of observed AGN fractions in galaxies, as well as
for the empirical determination of SMBH mass functions with large
surveys. We confirm previous
results and clearly demonstrate that, at least for $z\la 1.5$, SMBH mass
function evolves anti-hierarchically, i.e. the most massive holes grew
earlier and faster than less massive ones. For the first time, we find
hints of a reversal of such a downsizing behaviour at redshifts above
the peak of 
the black hole accretion rate density ($z\approx 2$). We also derive tight
constraints on the (mass weighted) average radiative efficiency of
AGN: under the simplifying assumption that the mass
density of both high redshift ($z\sim 5$) 
and ``wandering'' black holes ejected from
galactic nuclei after merger events are negligible compared to the
local mass density, we find that $0.065<\xi_0\langle \epsilon_{\rm
  rad} \rangle < 0.07$, where $\xi_{0}$ is the local SMBH mass density
in units of $4.3\times 10^5 M_{\odot}$ Mpc$^{-3}$. We trace the
cosmological evolution of the kinetic luminosity function of AGN, and
find that the overall 
efficiency of SMBH in converting accreted rest mass energy into
kinetic power, $\epsilon_{\rm kin}$, ranges between $\epsilon_{\rm
  kin} \simeq 3 \div 5 \times 10^{-3}$, depending on
the choice of the radio core luminosity function. Such a ``kinetic efficiency''
varies however strongly with SMBH mass and redshift, being maximal for very
massive holes at late times, as required for the AGN feedback by many
galaxy formation models in Cosmological contexts.

\end{abstract}

\begin{keywords}
accretion, accretion disks -- black hole
physics -- galaxies: active -- galaxies: evolution -- quasars: general
\end{keywords}

\section{Introduction}
\label{sec:intro}
Soon after their discovery, it was realized that Quasars (QSOs) were
a strongly evolving Cosmological population, that could effectively be
used as a tracer to study the structural properties of the evolving
Universe \cite{longair:66}. The physical understanding of QSOs and of their
lower-luminosity counterparts (both generally 
called Active Galactic Nuclei, AGN) as accreting supermassive black
holes (SMBH) immediately led to speculations about the presence of
their dormant relics in the nuclei of nearby galaxies, with Soltan
(1982) first proposing a method to estimate the mass budget of SMBH
based on the demographics and evolutionary paths of observed QSO/AGN.

The search for the local QSO relics via the study of their dynamical
influence on the surrounding stars and gas carried out since the launch of
the {\it Hubble Space Telescope} (see Richstone et al. 1998;
Ferrarese et al. 2008, and reference therein)
led ultimately to a shift of paradigm
in our understanding of the role of black holes in the formation of
structures in the Universe. The discovery of tight scaling relations
between SMBH masses and properties of the host galaxies' bulges
\cite{magorrian:98,gebhardt:00,ferrarese:00,tremaine:02,marconi:03,haering:04,hopkins:07a}
clearly pointed to an early co-eval stage of SMBH and galaxy growth.

However, local scaling relations have proved themselves unable to
uniquely determine the physical nature of the SMBH-galaxy coupling,
and a large number of feedback models have been proposed which can
reasonably well reproduce these relations
\cite{silk:98,fabian:99,kauffmann:00,umemura:01,cavaliere:02,wyithe:03,granato:04,dimatteo:05,king:05,sazonov:05,escala:06}. Most
likely, it will only be by
studying in detail the redshift evolution of these scaling relations
that we will gain critical insight on the nature of the interaction between growing black holes
and their host galaxies and on the nature of the accretion process and
of jet formation in various regimes. In fact, many have recently attempted to
push the study of scaling relations to $z>0$, despite the large
observational biases
\cite{merlonietal:04,mclure:06,shields:06,peng:06,wu:07,treu:07,salviander:07,shen:08,schramm:08},
but the question of if, and, in case, by how much did these relations
evolve with redshift has so far eluded any unambiguous answer. 

For all these reasons, it seems to us necessary to make the best
effort possible to
reconstruct the history of SMBH accretion in order to follow closely 
the evolution of the black hole mass function. 
That will serve as a benchmark to test
various models for SMBH cosmological growth as well as those for the black
hole-galaxy co-evolution. 

As opposed to the case of galaxies, where the direct relationship
between the evolving mass functions of the various morphological types
 and the distribution of star forming galaxies is not
straightforward due to the never-ending morphological and photometric
transformation of the different populations, the situation is much simpler
in the case of SMBH. For the latter we can 
assume their evolution is governed by a simple
continuity equation \cite{cavaliere:71,small:92,yu:02}, where the mass
function of SMBH at any given time can be used to predict that at any
other time, provided the distribution of accretion rates as a function
of black hole mass is known (see section~\ref{sec:methods} for more
details). An analogous continuity equation could in principle be
solved for the evolution of black hole spin (and thus for the overall
accretion efficiency), whose time derivative, however,
depends both on the accretion rate and on the merger rate evolution
(Berti and Volonteri 2008; King, Pringle and Hofmann 2008).

Consequently, the simple physical and mathematical 
structure of a black hole
implies that their intrinsic properties are essentially determined by
three physical quantities only: 
mass, spin and accretion rate. Leaving aside a few open issues
related to the physics of strongly super-Eddington flows,
accretion theory is a mature enough field to allow a relatively
robust mapping of observables (spectral energy distributions,
variability patterns, etc.) into these physical quantities associated to
SMBH growth. This reduces the problem of unveiling the census of
growing black holes to that of a wavelength-dependent assessment of
selection biases, bolometric corrections, and accurate determinations
of the distribution of extrinsic observational properties, the level
of intervening obscuration being the main one.

The fact that obscuration is a crucial factor in determining the
observed properties of AGN was recognized early after the discovery of
the X-ray background (XRB) radiation \cite{setti:89}. 
Subsequent generation of
synthesis models of the XRB \cite{madau:93,comastri:95,wilman:99,gilli:01,gilli:07}, following closely on the
footsteps of increasingly larger and deeper surveys, have
progressively reduced the uncertainties on the absorbing column density
distribution, which, coupled to the observed X-ray luminosity
functions, provide now an almost complete census of the Compton thin
AGN (i.e. those obscured by columns $N_{\rm H}<\sigma_{\rm T}^{-1} \simeq
1.5 \times 10^{24}$ cm$^{-2}$, where $\sigma_{\rm T}$ is the Thomson
cross section). Indeed, this class of objects 
dominates the counts in the X-ray energy band where
almost the entire XRB radiation has been resolved into individual
sources \cite{hasinger:98,rosati:02,alexander:03,moretti:03,worsley:05}.
Accurate determinations of the XRB intensity 
and spectral shape (for the most recent results see Churazov et
al. 2007; Ajello et al. 2008), 
coupled with the resolution of this radiation into
individual sources \cite{brandt:05},
 allow very sensitive tests of how the AGN luminosity
and obscuration evolve with redshift (the so-called 'X-ray background
synthesis models' see e.g. Gilli et al. 2007, and references
therein). 
In much the same way, accurate determinations
of the local SMBH mass density and of the AGN (bolometric) luminosity functions,
coupled with accretion models that
specify how the observed AGN radiation translates into a black hole
growth rate, allow sensitive tests of how the SMBH population (its mass
function) evolves with redshift. This is the aim of this work,
which, by analogy, we name `AGN synthesis modelling'.

Reconstructing in detail the history of accretion onto supermassive
black holes is also of paramount importance for understanding the
effects that the energy released by growing black holes has on its
environment. As we mentioned before,
the various relations between black hole mass and the physical
properties of host galaxy's bulge
suggest that the growth of black holes and large scale structure is
intimately linked. Moreover, X-ray observations of galaxy clusters
directly show that
black holes deposit large amounts of energy into their environment in
response to radiative losses of the cluster gas
\cite{birzan:04,fabian:06,allen:06}. From these (and other) studies of
the cavities, bubbles and weak shocks
  generated by the radio emitting jets in the intra-cluster medium
  (ICM) it appears that AGN are
 energetically able to balance radiative losses from the ICM
 in the majority of cases \cite{rafferty:06}. 
Also, numerical simulations of AGN-induced feedback have recently shown
that mechanical feedback from black
holes may be responsible for halting star formation in
massive ellipticals, perhaps explaining the bimodality in the
color distribution of local galaxies \cite{springel:05}, as well as
the size of the most massive ellipticals \cite{delucia:07}.

All these arguments, however, hinge on the unknown efficiency 
with which growing black holes convert accreted rest mass into kinetic
($\epsilon_{\rm kin}$) and/or radiative ($\epsilon_{\rm rad}$)
power.  Constraints on these efficiency factors are therefore vital for all models
of black hole feedback, and we aim here at pinning them down as
accurately as currently possible.  

The structure of this paper is the following: we will first start with
a description of our current knowledge about the various modes in
which a black hole may grow (\S~\ref{sec:modes}). This will be used to
introduce the function that relates the observed (bolometric)
luminosity of an AGN to the true accretion rate. In
section~\ref{sec:methods} we will outline the methods used to
calculate the evolution of the SMBH mass (and accretion rate)
function, while in section~\ref{sec:assumptions} we will spell out and
discuss the fundamental assumptions made in this work, in particular
those about the local mass function of SMBH, the adopted bolometric
correction and the incidence of obscured AGN. Our results on the
growth of the SMBH population will be presented in
\S~\ref{sec:mf}. There, we will first discuss ``integral'' constraints
(\S~\ref{sec:integral}) and then ``differential'' ones
(\S~\ref{sec:mf_sub}-\ref{sec:active}), which 
will reveal interesting features of the so-called anti-hierarchical
(or 'downsizing') evolution of SMBH. Section~\ref{sec:kinetic} will
instead be devoted to the study of the kinetic luminosity function of
AGN and of its evolution. At the end of it (\S~\ref{sec:feedback}) we
will be able to assess the relative importance of kinetic and radiative
feedback from growing black holes as a function of both mass and
redshift. Finally, we will discuss our results and draw our
conclusions in section~\ref{sec:discussion}.

Throughout the paper we will use concordance
cosmological parameters of $\Omega_{\rm M}=0.3$,
$\Omega_{\Lambda}=0.7$, $H_{0}=70\,{\rm km\,s^{-1}\,Mpc^{-1}}$.

\section{The modes of black hole growth}
\label{sec:modes}

Galactic (stellar mass) black holes in X-ray binaries, 
either of transient or persistent
nature, are commonly observed undergoing so-called transitions,
i.e. dramatic changes in their spectral and variability properties
associated to changes in the way matter accretes onto them. The study
of these various spectral states, either in individual objects or in ensemble
of sources, has solidified in the last few years 
into a general framework of the various
physical modes in which accretion onto a black hole can take place
\cite{remillard:06,malzac:07,done:07}. We discuss these modes, and
their AGN analogues, in this
section, paying particular attention to their partition in terms of
the value of the Eddington ratio  $\lambda \equiv  L_{\rm bol}/L_{\rm
  Edd}$, where $L_{\rm bol}$ is the bolometric luminosity and $L_{\rm
  Edd}=4 \pi G M_{\rm BH}
m_{\rm p} c / \sigma_{\rm T} \simeq 1.3 \times 10^{38} (M_{\rm
  BH}/M_{\odot})$ erg s$^{-1}$ is the Eddington one.

There are at least three well defined  spectral states. In the {\it
  low/hard} state the emission is dominated by a hard X-ray power-law
 with an exponential cutoff at about few 100 keV. Radio emission is
 always detected in this state \cite{fender:06} usually with flat or
 inverted spectrum, which is commonly interpreted as due to a compact,
 persistent, self-absorbed jet of low intrinsic power. Interestingly,
 the radio luminosity of these jet  cores seems to correlate quite
 tightly with the X-ray luminosity \cite{gallo:03}.
The spectrum of the
 {\it high/soft } state, 
instead, is dominated by a thermal component 
likely originated in a standard Shakura \& Sunyaev (1973) accretion disc,
while in the {\it intermediate (or very high)} 
state, associated with transitions between hard and soft states, and
often occurring at a source's
highest flux level, both a thermal and a steep power-law 
component substantially contribute to the spectrum\footnote{In fact,
  the overall evolution of transient black holes does show clear
  evidences of hysteresis \cite{miyamoto:95,belloni:05},
  with the intermediate 
  spectral states having different luminosity when sources enter or
  leave the outburst. We neglect this further complication
  here.}. Maccarone (2003)  
has shown that the state change in these systems generally
occurs at (bolometric) luminosities of about $\lambda \simeq {\rm a \;
  few} \times 10^{-2}$. The hard-to-soft transition is also accompanied
by a ``quenching'' of the steady radio emission observed in the
low/hard state, following rapid flaring events in the radio
band, often associated with fast ejection of
matter moving at relativistic speeds and emitting optically thin
synchrotron radiation \cite{fender:04}. 

Given the many analogies between the high-energy spectra of
galactic black holes and AGN, a similar phenomenology has long been
searched for in active supermassive black holes. Merloni, Heinz and Di
Matteo (2003; MHD03), following earlier suggestions by Falcke and
Biermann (1996), presented a thorough
discussion on the scale-invariant properties of black hole coupled 
accretion/jet systems, based essentially on a generalization of the
radio-X-ray correlation of low/hard state X-ray binaries to SMBH of
varying masses. This led to the discovery of a
 so-called ``fundamental plane'' of active black
holes (see also Falcke, K{\"o}rding and Markoff 2004;
K{\"o}rding, Falcke and Corbel 2006; Merloni et al. 2006; Fender et
al. 2007).

This correspondence between low/hard state X-ray binaries and
low-luminosity AGN is confirmed by a number of recent studies. 
Detailed multi-wavelength observations of nearby galaxies
 have revealed a clear tendency for lower
luminosity AGN to be more radio loud as the Eddington-scaled accretion
rate decreases (see Ho 2002, and references
therein; Nagar, Falcke and Wilson 2005).
The above mentioned scaling relations with black hole X-ray binaries 
 have thus helped to identify AGN analogues
of low/hard state sources, in which the radiative efficiency of the
accretion flow must be low
\cite{churazov:05,pellegrini:05a,chiaberge:05,hardcastle:07}. 
 
In this work, we will use the kinetic jet power as a key physical
variable that identifies and characterizes the nature of various
accretion modes. In fact, Merloni and Heinz (2007), using a
sample of nearby AGN for which BH masses, nuclear radiative and
kinetic luminosities and outer boundary condition (i.e. accretion rate
at the Bondi radius) were simultaneously known, have found 
a clear relationship between Eddington-scaled kinetic power 
and bolometric luminosity, 
given by: $\log (W_{\rm kin}/L_{\rm Edd}) = (0.49\pm0.07) \log \lambda - (0.78\pm0.36)$. The measured
 slope suggests that these objects are in a
radiatively inefficient accretion mode. Moreover, 
the observed correlations are in very good
agreement with theoretical predictions of adiabatic accretion
models with strong outflows, the relative strength of which increases
with decreasing accretion rate \cite{blandford:99,blandford:04}. 

This is a very important finding, as the existence of jet dominated
accreting SMBH bears important consequences for our understanding of
the AGN feedback in a cosmological context.
The specific importance of
mechanical (as opposed to radiative) feedback 
for the heating of baryons within the deepest dark
matter potential wells has recently been acknowledged by semi-analytic
modelers of cosmological structure formation
\cite{croton:06,bower:06}.
Within these schemes, because the bright quasar population peaks at
too early times, a so-called ``radio mode'' of SMBH growth is invoked
in order to regulate both cooling flows in galaxy clusters
 and the observed sizes and colors of
the most massive galaxies in the local Universe \cite{springel:05,croton:06}. 

Here, we also consider such a mode of accretion, where most of the
released energy is in kinetic form, but we uniquely associate it
 to black holes accreting below a certain
critical rate in Eddington units, rather than to the state of the
fuelling gas \cite{hardcastle:07}, or to the properties of the very
large scale environment (as, for example, the host dark matter halo size). 
In order to avoid confusion,
we call this ``low kinetic (LK)'' mode, to distinguish it not
only from the bright ``high radiative (HR)'' mode (the 'quasar' mode of
the recent cosmology jargon, associated to radio quiet QSOs, analogous
to the high/soft state of X-ray binaries), 
but also from the bright radio loud
quasars (of which 3C273 is the prototype), characterized by both 
radiatively efficient accretion flows and powerful jets, that we term
``high kinetic (HK)'' mode, as we discuss below. 

The most luminous sources, as those falling into the
standard definition of Quasars and broad lined AGN,
 are most likely accreting at a high
rate, close to the Eddington limit \cite{mclure:04,kollmeier:06}. 
Also their spectral energy
distribution, usually dominated by the so-called Big Blue Bump
(BBB: quasi thermal UV emission from an optically thick standard accretion
disc, see e.g. Malkan 1983; Laor 1990), 
indicates that above a certain critical value of the ratio
$\lambda_{\rm cr}$, an accretion mode transition should take place  to what
is usually described as a standard, geometrically thin and optically
thick accretion disc (Shakura \& Sunyaev 1973). Evidence for such a
transition taking place in AGN has
been presented by  Maccarone, Gallo \& Fender (2003), Jester (2005)
and  Greene, Ho
and Ulvestad (2006).

Thus, according to our scheme, black holes accreting {\it above}
the critical rate come into two different physical states, one with
(HK), and one without powerful jets (HR). The reason for this dichotomy
(and even its reality, see e.g. Cirasuolo et al. 2003) 
has been subject to countless arguments since
the very early years of the first radio loud QSOs discoveries, but it
is still matter of debate (for example, see Sikora, Stawarz \& Lasota
2007; Kaiser \& Best 2008; Blundell 2008 for some recent ideas), with the two most
popular options for the observed dichotomy
being the influence of the black hole spin and a
change in accretion mode. 

As a working hypothesis,
in line with what originally proposed in MHD03 (see also Nipoti et
al. 2005; K{\"o}rding,
Jester and Fender 2006; Blundell 2008), we consider here that
powerful radio jets are episodic and transients events during the time
a black hole spends accreting above the critical rate (i.e. close to
the Eddington limit). This idea, that is easy to incorporate into any
population synthesis scheme, bears resemblance with what actually
observed for transients black hole X-ray binaries, which often
displays very powerful and rapid radio flares at the peak of their
outbursts (see Fender et al. 2004, and references therein).

\subsection{A schematic view of radiative, kinetic and accretion efficiency}
\begin{figure}
\psfig{figure=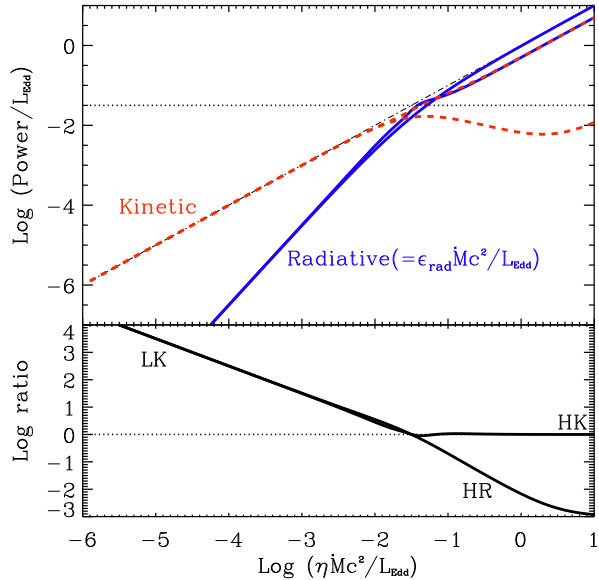,height=8.5cm}
\caption{A schematic view of the relationship between accretion rate
  and released power, both in units of the Eddington luminosity 
(an ``accretion mode map''). Blue solid lines
  show the radiated power, red dashed ones the released kinetic
  power. The black horizontal dotted line mark the critical Eddington
  ratio where a change of accretion mode is assumed to 
occur. Below this rate, only
one mode (LK) is allowed, where most of the power emerges in kinetic
form; above it, two different modes are possible: one where 
kinetic and radiative power are comparable and high (HK), and one
where kinetic power is quenched (HR). The lower panel shows the
corresponding ratios
between kinetic and radiative power as a function of accretion rate.}
\label{fig:modes}
\end{figure}

We adopt a simple, physically motivated
functional form for the $\lambda$ vs. $\dot m$ function (where $\dot m$ is the
accretion rate onto the black hole in Eddington units $\dot m = \eta
\dot M c^2/L_{\rm Edd}$, with $\eta$ accretion efficiency) of a broken
power-law, bridging the low accretion rate (radiatively inefficient)
regime ($\lambda \propto \dot m^2$) and the high
accretion rate one. For radiatively efficient sources, we should
assume, by definition, that the bolometric luminosity is simply
proportional to the accretion rate, $\lambda \propto \dot m$. 
The overall
normalization is found by imposing continuity at $\lambda_{\rm cr}$ and 
depends on the adopted value of this critical accretion rate. Based on the
results of Maccarone et al. (2003), Greene et al. (2006), Merloni and
Heinz (2007), we adopt here a value of $\lambda_{\rm cr}=3\times
10^{-2}$ (see also the discussion in Hopkins, Narayan \& Hernquist
2006). A schematic view of these various modes in the accretion rate
vs. released power plane is shown in Fig.~\ref{fig:modes}. The Figure
represents qualitatively the main features of the three-modes scheme
we have outlined here: below the critical rate, only
one mode (LK) is allowed, where most of the power emerges in kinetic
form and there is a quadratic relationship between accretion rate and
(bolometric) radiative luminosity; 
above it, two different modes are possible: one where 
kinetic and radiative power are comparable and high (HK), and one
where kinetic power is quenched (HR). The
exact slopes of the kinetic power vs. accretion rate relation are
set, implicitly, by the relationship between radio core luminosity,
kinetic power and X-ray luminosity (the ``fundamental plane'' of active
black holes), as well as by the hard X-ray bolometric correction
function, that we discuss in \S~\ref{sec:methods} and \ref{sec:assumptions}.

This simple relation, together with a bolometric correction,
 allows the determination of the dimensionless
accretion rate for any object for which both the nuclear luminosity, $L_i$,
(in any band) and the black hole mass are measured. Conversely, the
accretion luminosity of a black hole of mass $M$, growing at a rate
$\dot M$ can be obtained from the above relationship. In 
\S~\ref{sec:methods} we will describe in detail the method we employ
to unveil the evolution of the black hole mass function based on the
evolution of the intrinsic hard X-ray (2-10 keV) luminosity function,
making use of the relation $L_i=L_i(\dot M,M)$ discussed above. 

As a final remark in this section, we come back here to the well known distinction
between {\it accretion efficiency}, and {\it radiative efficiency},
that will be important in the following. The
former represents the maximal amount of
potential energy that can be extracted, per unit rest mass energy,
from matter accreting onto the black hole. This quantity, $\eta(a)$,
depends on the inner boundary condition of the accretion flow only,
and, in the classical no-torque case \cite{novikov:73} is a
function of BH spin, $a$, only, ranging from $\eta(a=0)\simeq 0.057$ for
Schwarzschild (non-spinning) black holes to $\eta(a=1)\simeq 0.42$ for
maximally rotating Kerr black holes. 

On the other hand, the radiative efficiency, $\epsilon_{\rm rad}\equiv
L_{\rm bol}/\dot M c^2$ depends both on the accretion efficiency
(i.e. on the inner boundary condition of the accretion flow) and on
the nature of the accretion flow itself. In the case of a simple broken
power-law shape that we have assumed for the $\lambda-\dot m$ relation, we
have that: 
\begin{equation}
\label{eq:radeff}
\epsilon_{\rm rad} \equiv \eta f(\dot m) =\eta \times \left\{
        \begin{array}{ll}
        1,   &  \dot m \ge \dot m_{\rm cr}   \\
        (\dot m/\dot m_{\rm cr}), & 
\dot m < \dot m_{\rm cr}   \\
        \end{array}\right.\;
\end{equation}
where 
$\dot m_{\rm cr}=\lambda_{\rm cr}$ is the
critical Eddington-scaled accretion rate
above which the disc becomes 
radiatively efficient. As we will show in \S~\ref{sec:mf}, 
the classical argument
due to Soltan (1982) can be used to put constraints on the (mass
weighted) average value of $\langle \epsilon_{\rm
  rad}\rangle$, by comparing the evolution of AGN/QSO luminosity function with
estimates of the local relic black hole mass functions. Constraints on
$\langle\eta(a)\rangle$, and thus on the average spin of the SMBH
population,
 can then be derived from the use of
eq.~(\ref{eq:radeff}) \cite{hopkins:06a}. The
relationship between these average values $\langle\eta\rangle$ and
$\langle\epsilon_{\rm rad}\rangle$ depends both on the value of $\lambda_{\rm
  cr}$ (higher values of the critical accretion rate increase the
number of objects which are in a radiatively inefficient regime, thus
reducing the ratio between radiative and accretion efficiencies, and
viceversa) and on the shape and evolution of the luminosity functions
themselves. However, in the remain of the paper we will assume that
eq.~(\ref{eq:radeff}) holds, and will not discuss explicitly the
dependence of the $\langle \eta\rangle /\langle \epsilon_{\rm
  rad}\rangle$ ratio on $\lambda_{\rm cr}$. Moreover, for the sake of
clarity, we will refer in the following to the mass weighted averages
of the efficiencies simply with the symbols $\eta$ and $\epsilon_{\rm rad}$.

\section{Methods}
\label{sec:methods}
In the simplest case of 
purely accretion driven evolution (i.e. assuming mergers do not play
an important role in shaping the black holes mass function, but see
\S~\ref{sec:discussion} for more on this point), the
simultaneous knowledge of the black hole mass function and of the
 {\it Eddington ratio
distribution function}, $\Phi_{M,\lambda}(z)$ (i.e. the number of
objects with given mass $M$ and Eddington ratio $\lambda$ 
per comoving volume and logarithmic interval),
 would completely determine the evolutionary
solution for the mass function, as  the two distributions must be coupled via a
continuity equation (Small \& Blandford 1992; Yu \& Tremaine 2002; 
Marconi et al. 2004;
Merloni 2004, hereafter M04; Shankar, Weinberg \& Miralda-Escud{\'e} 2008)
\begin{equation}
\label{eq:continuity}
\frac{\partial \Phi_{\rm M}(M,t)}{\partial t}+\frac{\partial
[\Phi_{\rm M}(M,t) \cdot \langle\dot M(M,t)\rangle]}{\partial M}=0,
\end{equation}
to be solved given an appropriate boundary condition. 
Here $\Phi_{\rm M}$ is the black hole mass function, 
and $\langle \dot M(M,t)\rangle$ represents the average accretion rate
(in physical units) 
for a SMBH of mass $M$ at the time $t$, and can be uniquely
determined once the  black hole Eddington ratio
distribution function, $\Phi_{M,\lambda}(z)$, is known.

 The method used to solve eq.~(\ref{eq:continuity})
follows closely that 
applied in M04, with some modifications and improvements,
as we describe it in detail here.
As in M04, we argue that the most robust constraint on
the SMBH mass function at any redshift can be obtained by integrating the
continuity equation {\it backwards} in time, using the local mass
function as a boundary condition. Moreover, as opposed to all other 
previous calculation of the SMBH mass function
evolution, we do not make any
simplifying assumption about the distribution of the accretion rates
for different black hole masses and redshifts, but we do 
calculate these distributions from the multi-wavelength luminosity 
functions of the AGN population.
The guiding principle is that, within each of the three 
physically different modes of accretion described in the previous
section (LK, HR, HK), there exist well defined relationships between
an object luminosity in various bands and the physical parameters
needed to solve eq.~(\ref{eq:continuity}), i.e. the black hole mass and
its accretion rate. In section~\ref{sec:modes} we have already presented
the relationship between accretion rate, mass and bolometric
luminosity in these various modes. The second relationship, needed in
order to determine the evolution of 
the AGN jet (and thus of their kinetic energy feedback) is that between radio
luminosity, bolometric luminosity and BH mass, that we discuss here. 

In the 'low kinetic' (LK)
mode, i.e. for all sources with Eddington ratios $\lambda$ less than
the critical value $\lambda_{\rm cr}$, we adopt the now well
established result, first described in MHD03, 
that a 'fundamental plane' relation between (core radio) jet
luminosity, (accretion-powered) X-ray luminosity and black hole mass
can indeed be defined for those objects (but see, for example, Panessa
et al. 2007 for a contrasting view). 
We write it here as 
\begin{equation}
\label{eq:fp}
\log L_{\rm R}=\xi_{\rm RX}\log
L_{\rm X}+\xi_{\rm RM} \log M + \beta_{\rm LK}, 
\end{equation}
where $M$ is the black hole mass in solar units, 
$L_{\rm R}$ is the {\it intrinsic} 
(i.e. un-beamed) radio luminosity of the jet core
at 5 GHz and $L_{\rm X}$ stands for the {\it intrinsic} (i.e. unabsorbed)
X-ray luminosity in the 2-10 keV band (both in units of erg s$^{-1}$), which we uniquely relate to
the bolometric luminosity by the relation of eq. (21) of Marconi et
al. (2004). We adopt for the correlation coefficients
the values recently determined in Merloni and Heinz (2007)
based on a more accurate study of the effects of relativistic beaming,
as well as a more accurate selection of the parent sample, as first
discussed in K{\"o}rding, Falcke and Corbel (2006). Specifically, we
assume $\xi_{\rm RX}=0.62$, $\xi_{\rm RM}=0.55$ and $\beta_{\rm
  LK}=8.6$. We also assume that this relationship has an intrinsic
scatter of 0.6 dex, a value smaller than that found in MHD03, but in
line with that discussed in K{\"o}rding et al. (2006) for samples of
jet dominated objects only.

Let us now discuss the relationship between radio, X-ray luminosity
and black hole mass for objects accreting at high rates.
Irrespective of the physical explanation preferred for the existence
of a radio loud/quiet (HK/HR) dichotomy, the problem we face for our
method is that there are no well
defined relationships to be found in the literature between radio
luminosity, mass and bolometric luminosity which are specific for
black holes accreting {\it above} the critical rate, $\lambda_{\rm
cr}$. 
One reason for this, as
we have argued before in Merloni et al. (2005), is that, because such modes
can exist only in very limited range of the parameter space (i.e. for
$\lambda_{\rm cr}\la \lambda \la 1$, where $\lambda=1$ corresponds to
the Eddington limit), it is almost impossible to unravel the mass from
the accretion rate dependence, given the usual uncertainties on SMBH
mass estimates and the likely presence of intrinsic scatter in any
such relation.
Thus, our treatment of the HK sources (classical radio loud QSOs) should
necessarily be considered 
at best as an educated guess. For the sake of simplicity, 
we assume that their intrinsic radio core luminosity scales with the bolometric
luminosity, so we fix 
\begin{equation}
\label{eq:fp_h}
\log L_{\rm R}=\log L_{\rm bol}+\xi_{\rm HM}
\log M + \beta_{\rm HK,HR}. 
\end{equation}
We work out the mass scaling $\xi_{\rm
  HM}$ (the same for HK and HR modes) and the normalization of the HK
(radio loud) mode assuming continuity with the fundamental plane
relation at the critical accretion rate, while for the HR (radio
quiet) mode, we simply take a luminosity  10$^3$ times smaller
($\beta_{\rm HR}=\beta_{\rm HK}-3$), so that the radio luminosity in
the HR (radio quiet) mode is effectively quenched. We obtain
$\xi_{\rm HM}=0.02$ (mass dependency is negligible), and
$\beta_{\rm HK}=-4.9$, in relatively good agreement with the standard
radio loud SED of Elvis et al. (1994). Also for this relationship we
assume an intrinsic scatter of 0.6 dex.

Once fixed the relationship between $L_{\rm R}$, $L_{\rm X}$ and $M$ for all
values of the accretion rate (i.e. for the three main modes LK, HR and
HK), we can proceed in extracting the
distribution of accretion rate as a function of SMBH mass and
redshift. In order to do so, we take the observed luminosity functions
for $L_{\rm R}$ (intrinsic radio core emission) and $L_{\rm X}$ 
(intrinsic 2-10 keV luminosity) and
 the SMBH mass function. We then look for the joint distribution
$\Psi_C(L_{\rm R},L_{\rm X},M)$, i.e. the function that quantifies the
differential number of objects (per unit comoving volume) with given
mass $M$, radio luminosity $L_{\rm R}$ and X-ray luminosity $L_{\rm X}$.
We do so by imposing the relations (\ref{eq:fp}) and (\ref{eq:fp_h}),
with their own intrinsic scatter, and minimizing the differences
between the projections of $\Psi_C(L_{\rm R},L_{\rm X},M)$
onto the three axes identified by $L_{\rm R}$, $L_{\rm X}$ and $M$ 
and the observed luminosity and mass functions. For high accretion
rate objects (i.e. those with $\lambda>\lambda_{\rm cr}$), we assume
that the HK (radio loud) objects represent only 10\% of the total, so
that the number ratio of HR/HK AGN is about 9/1.
We also impose the Eddington limit by assuming that the number of objects with
$\lambda>1$ are exponentially suppressed. By construction, then, our
joint distribution function  $\Psi_C(L_{\rm R},L_{\rm X},M)$
reproduces the observed luminosity (in the radio and X-ray bands) 
and mass functions, and is consistent with the scaling relations
(\ref{eq:fp}) and (\ref{eq:fp_h}). 


Once the joint distribution $\Psi_C$ is found, it is straightforward
to infer the accretion rate distribution function: each combination
($L_{\rm R}$, $L_{\rm X}$ and $M$) correspond to a unique value of
the accretion rate $\dot M$, and the number of objects per unit
comoving volume (per logarithmic interval) 
with a given value of $M$ and $\dot M$ is given by:
\begin{eqnarray}
\label{eq:joint_mdot}
\Phi_{M,\dot M} &\equiv& \frac{dN}{d\log M d\log \dot M} \nonumber \\ 
&=& \int
\Psi_C[L_{\rm R},L_{\rm X}(M,\dot M),M] \frac{\partial
  \log L_{\rm X}}{\partial \log \dot M} d\log L_{\rm R}
\end{eqnarray}
where $L_{\rm X}=L_{\rm X}(M,\dot M)$ is the relation between intrinsic X-ray
luminosity and mass and accretion rate discussed in section~\ref{sec:modes}.
Thus, the knowledge of the joint distribution function $\Psi_C$,
complemented with the theoretical relationship between luminosity and
accretion rate for the various modes ($X(M, \dot M)$, see \S~\ref{sec:modes}) is
sufficient to derive the distribution of accretion rates for black
holes of different masses. The average $\langle \dot M(M)\rangle$ is
what is then needed to solve eq.~(\ref{eq:continuity}).

In this way, the combined knowledge of the SMBH mass
function and of the AGN luminosity function in at least 
one band at a given moment in time 
completely determine the boundary conditions needed to integrate 
the continuity equation.
In practice, starting from the mass function and the accretion rate distribution
 at a given redshift $z_i$, 
it is possible to derive the new black hole mass function at redshift
 $z_i+dz$, 
$\Phi_{\rm M}(M, z_i+dz)$, by just subtracting the mass accreted in the
 time interval $dt=dz(dt/dz)$ calculated according to the accretion
 rate function of redshift $z_i$. This new mass function can then be
used together with the radio and X-ray luminosity functions at the
 same redshift,
$\phi_{\rm R}(L_{\rm R}, z+dz)$ and $\phi_{\rm X}(L_{\rm X},
z+dz)$, to obtain the new joint distribution function 
$\Psi_{C}(L_{\rm R},L_{\rm X},M,z+dz)$, 
and therefore the new accretion rate function,
and so on, until a solution of eq.~(\ref{eq:continuity}) is found in
the interval $z_i+\Delta z$, where the value of $\Delta z$ depends
solely on the availability of good enough data (luminosity functions)
to meaningfully determine the joint distribution $\Psi_C$. 


\section{Assumptions and observables: mass and luminosity functions}  
\label{sec:assumptions}
According to the procedure we have outlined in \S~\ref{sec:methods} above, 
in order to obtain meaningful constraints on the accretion rate
distribution, we have to make the following assumptions:
\begin{itemize}
\item{The local BH mass function is known accurately down to a certain
SMBH mass;}
\item{The bolometric corrections are known at all redshifts for AGN of all
luminosities, so that meaningful bolometric luminosity functions can
be derived;}
\item{The selection function of hard X-ray surveys provides us with
a complete census of the AGN population, i.e. is not strongly
biased against obscured sources. Although this is obviously not the
case for the so-called Compton Thick AGN, i.e. those obscured by such
a large column ($N_{\rm H}>\sigma_{\rm T}^{-1}$) that also hard X-rays cannot be transmitted to the
observer, we will proceed under the assumption that the Compton Thick
AGN fraction can be meaningfully constrained on the basis of the X-ray
background models, and verify, {\it a posteriori} that their
contribution to the overall growth is small;}
\item{The bolometric LF of AGN, once extrapolated to low enough luminosities,
recovers the entirety of the local SMBH population, as described by
the adopted mass function. This is equivalent to the statement that
the AGN fraction approaches unity towards very low nuclear
luminosities, once incompleteness of any
specific LF is accounted for.}
\end{itemize}
We will discuss these issues in turn below.

\subsection{Local black hole mass function}
\label{sec:local}
 As it is well known, uncertainty in the local mass function
 determination propagates linearly in the computation of the
average accretion efficiency in Soltan-type of arguments.
This issue has been dealt with extensively in the
literature  \cite{soltan:82,marconi:04,merlonietal:04}. Recently, progress have
been made towards a more secure computation of the local mass function
of supermassive black holes, based on the various scaling
relations. For example, Graham and Driver (2007) have shown how
most previous inconsistencies among local black hole mass densities,
$\rho_{\rm BH,0}$, computed using different scaling relations were due
to incorrect accountancy for the Hubble constant dependence. 
Here, we will not dwell into a lengthy discussion about the
available constraints on the local distribution of SMBH masses,
for which we refer the reader to the comprehensive treatment presented
 in Shankar et al. (2008b), who 
have carried out a detailed comparison among SMBH mass
functions derived with different methods. Rather, we briefly
describe in the following our choice of the initial condition for
eq.~(\ref{eq:continuity}). Although uncertainties do
remain, with the local black hole mass density value estimated within
the range $\rho_{\rm BH,0}=(3.2 - 5.4) \times 10^{5}$ M$_{\odot}$
Mpc$^{-3}$ (for $H_{0}=70$ km s$^{-1}$ Mpc$^{-1}$), the agreement
between various measurements is encouraging. We will adopt a
specific analytic expression for the SMBH mass function, 
coincident with the central value within the
uncertainty range in Shankar et al. (2008b). Our adopted mass function
is computed as the convolution of a Schechter function
$\Psi_{\rm M}$ with a Gaussian scatter of 0.3 dex, to account for the
intrinsic scatter of the scaling relations from which the mass
function is derived.
For the Schechter function, we adopt the following parametrization
\begin{equation}
\Psi_{\rm M}=\phi_*\left(\frac{M}{M_{*}}\right)^{1+\alpha}\exp\left({1-\frac{M}{M_{*}}}\right)
\end{equation}
with parameters: $\phi_*=10^{-3}$, $\log M_{*}=8.4$ and
$\alpha=-1.19$. With this choice, the local black hole mass density is 
$\rho_{\rm BH,0}=4.3 \times 10^{5}$ M$_{\odot}$
Mpc$^{-3}$ and
the total number density of black holes above $5\times 10^5$ solar
masses is about $n_{\rm BH}\simeq 1.3 \times 10^{-2}$ Mpc$^{-3}$.

\begin{figure}
\centering
\psfig{figure=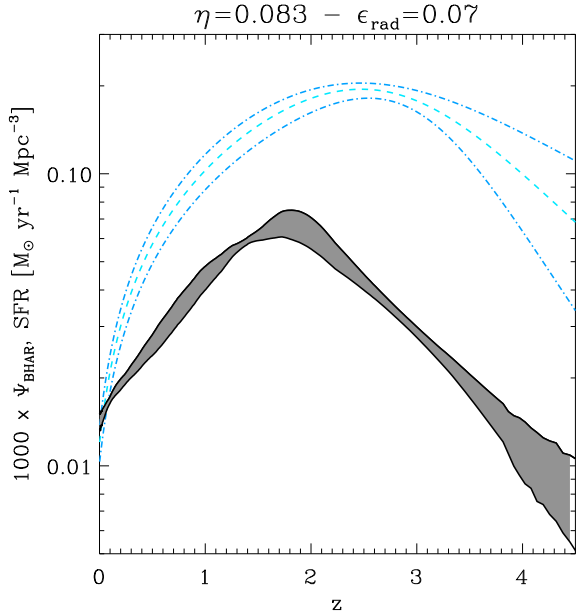,height=8.5cm}
\caption{Redshift evolution of the accretion rate density onto
  supermassive black holes
$\Psi_{\rm BHAR}(z)$, multiplied by a constant (equal to 10$^3$) is
plotted as a shaded area alongside the best fit to a large compilation
of data for the star formation
rate (SFR) density (blue dashed, from Hopkins and Beacom 2006). The shaded area
represents the uncertainty in the black hole accretion rate density
deriving from the observational uncertainty on the AGN
luminosity functions.}
\label{fig:zev_mdot_datasfr}
\end{figure}

\subsection{Bolometric corrections and luminosity functions}
\label{sec:bol_lf}
The cosmic evolution of the SMBH accretion rate and its
 associated mass density can be calculated from the bolometric luminosity
 function of AGN: $\phi(L_{\rm bol},z)$, where $L_{\rm
 bol}=\epsilon_{\rm rad} \dot M c^2$ is the (intrinsic, i.e. before
 absorption) bolometric luminosity produced by a
 SMBH accreting at a rate of $\dot M$ with a radiative efficiency
 $\epsilon_{\rm rad}$. In practice, the accreting black hole population is
 always selected through observations in specific wavebands.  Crucial
 is therefore the knowledge of two factors: the completeness of any
 specific AGN survey, and the bolometric correction needed in order to
 estimate $L_{\rm bol}$ from the observed luminosity in any specific
 band.

As for the accuracy of our knowledge of the bolometric correction, we
refer the reader to the studies of Marconi et al. (2004); Richards et al. (2006);
Hopkins et al. (2007b)
and Shankar et al. (2008b). All of them consistently demonstrate that a
luminosity dependent bolometric correction is required in order to
match type I (unabsorbed) AGN luminosity functions obtained by
selecting objects in different bands\footnote{This conclusion, however, has
been recently challenged by Vasudevan \& Fabian (2007), who ague that
the 2-10 keV X-ray bolometric correction is better correlated with
Eddington ratio rather than with luminosity.}. No significant trend with
redshift is instead evident, similarly to what is known for the
spectral properties of type I AGN in the X-ray band alone
\cite{vignali:03,tozzi:06,steffen:06}.
Here we will adopt, as fiducial inputs, the bolometric
corrections of Marconi et al. (2004) and the latest observed 2-10 keV luminosity
function of Silverman et al. (2008), which provides the best available
constraint on X-ray selected AGN at high redshift. In order to account for the
measurement uncertainties in a simple and straightforward way, we 
will use two different
analytical parametrization of the observed 2-10 keV luminosity function, as
discussed in Silverman et al. (2008): a luminosity dependent density
evolution model (LDDE; Ueda et al. 2003; Hasinger et al. 2005) and a
modified-pure luminosity evolution (MPLE; Hopkins et al. 2007b). The
analytic expressions for these models can be found in eqs.~(5-11) and
in Table 4 of Silverman et al. (2008).

\subsection{Accounting for obscured sources}
\label{sec:obscuration}
Hard X-ray selected samples of AGN represent the most unbiased census
of accretion activity in the Universe, due to the limited effect that
(Compton thin) gas obscuration has on the emerging luminosity in this
band. However, observed 2-10 keV luminosities of AGN are still
affected by absorption, and we need to quantify this effect when
looking for the intrinsic bolometric luminosity.

Assuming an average X-ray spectral model of the Gilli et
al. (2007) X-ray background synthesis model, with fixed
power-law slope of $\Gamma=1.9$, 
we can compute
the ratio of emerging to intrinsic 2-10 keV luminosity, $g\equiv
L_{\rm X,obs}/L_{\rm X}$, as a function of the
absorbing column density of cold gas, $N_{\rm H}$ (in
cm$^{-2}$). Using {\tt XSPEC} \cite{arnaud:96} we have computed the effect of
various obscuring columns on the emerging spectrum and then fitted 
the observed relation between $g$ and $N_{\rm H}$
with a fourth order polynomial in ${\cal{N}_{\rm
H}}=\log N_{\rm H}-21$, and found the following analytical expression,
accurate to better than 20\% in the whole range $10^{21}<N_{\rm
H}<10^{24}$:

\begin{equation}
g({\cal{N}_{\rm H}})\simeq 0.99+0.018{\cal{N}_{\rm H}}-0.059{\cal{N}_{\rm
H}}^2-0.1{\cal{N}_{\rm H}}^3+0.028{\cal{N}_{\rm H}}^4
\end{equation}

For Compton thick objects ($\log N_{\rm H}>24$), 
we have simply assumed that only a
scattered component of 0.02 times the intrinsic flux is detectable
(see Gilli et al. 2007).

Then, for any given observed 2-10 keV luminosity function
$\phi'(L_{\rm X})$, we can deconvolve the effect of obscuration if the 
distribution of column densities $f(N_{\rm H},L_{\rm X})$
as a function of luminosity is known. It is now well established that
the fraction of absorbed sources depends on the intrinsic X-ray
luminosity of an AGN
\cite{ueda:03,lafranca:05,treister:05,barger:05,hasinger:08}. 
Following again the latest
incarnation of the XRB synthesis model of Gilli et al. (2007), we can
write the ratio of absorbed to unabsorbed sources, $R(L_{\rm X})$ as
\begin{equation}
R(L_{\rm X})=1+2.7 e^{-L_{\rm X}/L_{c}}
\end{equation}
with $L_{c}=10^{43.5}$, so that the ratio of obscured to unobscured
sources tends to unity at high (QSO) luminosities and to 3.7 at low
luminosities. 

The distribution of absorbing columns is taken to be an increasing
function of the column density for Compton thin objects, and, for the
lack of better observational constraints, flat for Compton thick
sources, once again fully consistent with the XRB synthesis model of
Gilli et al. (2007). We thus define

\begin{equation}
f({\cal{N}_{\rm H}},L_{\rm X})=
\left\{ 
\begin{array}{ll}
A {\cal{N}_{\rm H}}\frac{R(L_{\rm X})}{1-R(L_{\rm X})}&   {\cal{N}_{\rm H}}\la3\nonumber \\
3 A \frac{R(L_{\rm X})}{1-R(L_{\rm X})} &   {\cal{N}_{\rm H}}>3 \\
\end{array} \right.
\end{equation}
where $A$ is fixed by the overall normalization $\int_0^{4.5}
f({\cal{N}_{\rm H}},L_{\rm X}) d{\cal{N}_{\rm H}} =1$.

Finally, we obtain the following expression for the intrinsic 2-10 keV
AGN luminosity function as a function of the observed (un-corrected) one:
\begin{equation}
\label{eq:xlf}
\phi(L_{\rm X}) \simeq \phi'(L_{\rm X}) - \frac{\partial
  \phi}{\partial \!\log L_{\rm X}}
\!\!\int_0^{4.5}\!\!\!\!\!\!\!
[1-g({\cal{N}_{\rm H}})]f({\cal{N}_{\rm
H}},L_{\rm X})d{\cal{N}_{\rm H}}
\end{equation}
and this is what we use in our calculation of $\Psi_C$ (section~\ref{sec:methods}).

\subsection{Do all SMBH have an ``active''  counterpart?}
\label{sec:doall}
Traditionally, the presence of an accreting supermassive black hole in
the nucleus of a galaxy have been associated to a number of
unambiguous signs of activity: high excitation broad and/or narrow
emission lines, 
hard X-ray luminosity exceeding 10$^{42}$ erg s$^{-1}$, the presence 
of relativistic jets, etc. Within this view, it is commonly held
belief that the fraction of AGN in
any randomly sampled population of galaxies is relatively
small. This is in obvious contrast with the idea that SMBH are
ubiquitous in the nuclei of galaxies (at least in the nearby
Universe), and have prompted many authors to search for the
``trigger'' that activate an otherwise quiescent black hole. 
However, it must be obvious that every observational definition of
AGN, like those mentioned above, suffers
from strong selection biases, in particular at low intrinsic
luminosities. It is in fact well known that high-resolution, high-sensitivity
studies of nearby galaxies not considered to be powerful AGN 
reveals that a very high fraction of them do indeed 
harbour SMBH showing clear signs of accretion. For example, Nagar et
al. (2005) and Filho et al. (2006) have used high-resolution radio imaging
of a magnitude limited sample of bright nearby galaxies (the Palomar
sample) to show that at least a quarter (and possibly as much as half)
of them are true AGN, albeit of very low luminosity. 
Similar results have been obtained by optical
spectroscopic studies \cite{ho:97} and by sensitive {\it Chandra}
X-ray observations \cite{ho:01,gallo:08,santra:07}.

Furthermore, we notice that at $z=0$, the low-luminosity slopes of both radio
\cite{nagar:05} and X-ray \cite{silverman:08} 
luminosity functions of AGN are steeper than the low mass
end slope of the SMBH mass function.
Therefore, if we limit ourselves to the study of
SMBH with (relic) masses larger than a given value, 
only a minor extrapolation of the observed luminosity functions towards low
luminosity is sufficient to reach total number densities comparable to that
of the local SMBH. 

For example, for our adopted luminosity functions, the number of SMBH with $M>5
\times 10^5 M_{\odot}$ is equal to the number of objects with {\it
  intrinsic} (i.e. un-absorbed) 2-10 keV luminosity larger than a few
$\times 10^{40}$ erg s$^{-1}$ and with {\it intrinsic}
(i.e. de-beamed, see \S~\ref{sec:fslf}) 5 GHz radio core luminosity
greater that a few $\times 10^{35}$ erg s$^{-1}$. From hard X-ray
surveys of the local Universe \cite{sazonov:07}, as well as from XRB
synthesis models \cite{gilli:07}, we know that at these very low
luminosities the number of X-ray AGN is dominated (in a ratio of
approximately 4:1) by heavily obscured objects, so we expect that
sensitive {\it Chandra} surveys reaching down to these luminosities
should unveil nuclear accreting black holes in a large fraction
of nearby galaxies, as indeed found by
Gallo et al. (2008). On the other hand, radio surveys are insensitive
to dust obscuration (although affected by relativistic beaming),
and indeed, based on the works of Nagar et al. (2005) and Filho et
al. (2006) it is reasonable to expect that essentially all nearby
galaxies will show compact, high brightness temperature cores with
$L_{\rm 5GHz}>10^{35}$ erg s$^{-1}$.

In our view, this justifies the assumption, implicit in all our
subsequent calculations, that there is no distinction between SMBH and
AGN, in the sense that every supermassive black hole is active, at some
level, like the SMBH at the center of our galaxy clearly demonstrates. 
Thus, the SMBH population coincides with that of all actively
accreting black holes, {\it once the broad distribution of accretion rates
is accounted for}. This is a very important point, at variance with most
current analytic, semi-analytic and numerical 
models of SMBH growth that often assume a fixed Eddington ratio, 
and the consequences
thereof will become manifest more than once in the remain of the
paper. In particular, we will discuss the observational implications for the
determination of AGN fraction in different surveys
in section~\ref{sec:active}.

\section{Unveiling the growth of supermassive black holes}
\label{sec:mf}
In this section, we concentrate on the growth of the
supermassive black hole population. We will proceed from the most
general to the more detailed results, namely from the study of integrated
quantities (SMBH mass and accretion rate density, radiative energy
density, average radiative efficiency, etc.) and their redshift
evolution, to differential ones,
focusing on the full mass and
accretion rate functions, and their evolution. The first part
(\S~\ref{sec:integral}) will
shed light on some very general properties of the cosmological growth
of SMBH, while the second (\S~\ref{sec:mf_sub}-\ref{sec:active}) 
will reveal interesting features of their so-called anti-hierarchical 
behaviour. 

In this section, as well as in the next one, all results will be shown
accounting for the intrinsic uncertainties of the adopted luminosity
functions. We estimate that these uncertainties can be evaluated by
comparing different analytic parametrization of the same data sets 
(the LDDE and MPLE parametrization for the hard X-ray luminosity function of
Silverman et al. 2008, \S~\ref{sec:bol_lf} and the two alternative
parametrizations for the flat-spectrum radio luminosity function of
Dunlop and Peacock (1990) and De Zotti et al. (2005), see \S~\ref{sec:kinetic}).
Then, we will solve the continuity equation for the
SMBH evolution according to the method described in section~\ref{sec:methods}
for the four possible combinations of X-ray and radio luminosity functions,
and present the results as shaded areas encompassing the entire range
spanned by these four solutions.
 
\begin{figure}
\centering
\psfig{figure=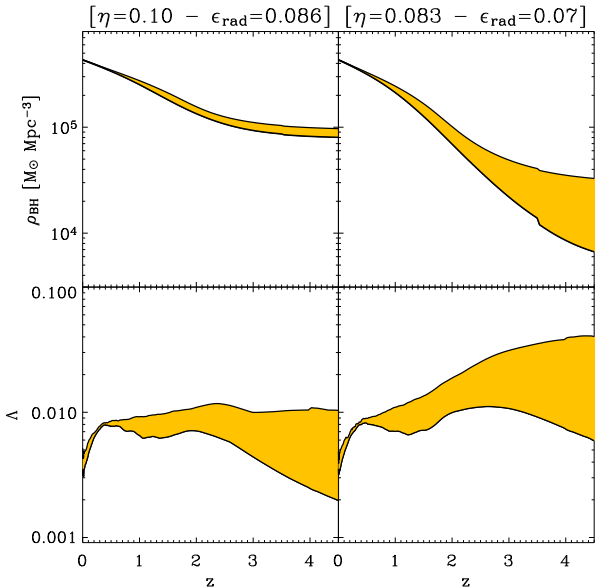,height=8.5cm}
\caption{Top panels: the redshift evolution of the SMBH
  mass density. Bottom panels, the evolution of the average Eddington
  ratio of the SMBH population, defined in eq.~(\ref{eq:aveledd}). On
  the left we show the results of a calculation performed fixing the
  accretion efficiency to $\eta=0.1$, which, for the particular model
  of accretion modes we assume correspond to an average radiative
  efficiency of $\epsilon_{\rm rad}\simeq 0.086$, while on the right
  we show the case of $\eta=0.083$, corresponding to $\epsilon_{\rm
    rad}\simeq 0.07$. See text for details.}
\label{fig:zev_100_083}
\end{figure}

\subsection{Integral constraints, mass density and radiative efficiency}
\label{sec:integral}
The redshift evolution of the black holes
accretion rate (BHAR) density can be easily calculated from the intrinsic
X-ray luminosity function (\ref{eq:xlf}) as follows:
\begin{equation}
\label{eq:bhar_z}
\Psi_{\rm BHAR}(z)=\int_0^{\infty}\frac{(1-\epsilon_{\rm rad})L_{\rm
    bol}(L_{\rm X})}{\epsilon_{\rm rad} c^2}
\phi(L_{\rm X},z)d\log L_{\rm X},
\end{equation}
where $L_{\rm X}$ is the intrinsic X-ray luminosity in the rest-frame 2-10 keV
band, and the bolometric correction function $L_{\rm bol}(L_{\rm X})$
is given by eq.~(21) of Marconi et al. (2004). 

In Figure~\ref{fig:zev_mdot_datasfr} we show the redshift evolution of
$\Psi_{\rm BHAR}(z)$, multiplied by a constant (equal to 10$^3$) in
order to visually compare its shape with that of the star formation
rate (SFR) density, which we take from the best fit to a large
compilation of data from Hopkins and Beacom (2006). The evolution
of the BHAR density of Fig.~(\ref{fig:zev_mdot_datasfr}) has been
computed for $\epsilon_{\rm rad}\simeq 0.07$, but changing this value
would simply result in a change of the overall normalization.

By construction, the computed SMBH accretion rate density 
includes both absorbed and unabsorbed
sources as well as Compton thick AGN (\S~\ref{sec:obscuration}), 
based on the prescription of
Gilli et al. (2007) that uses, as almost unique
observational constraint, the shape and normalization of the XRB
 radiation, and is thus somewhat degenerate in the
luminosity-redshift plane (see, however, Worsley et al. 2005). As a
sanity check, we can compute the
total contribution of Compton thick sources to the total mass
accumulated in SMBH today, which turns out to be
 limited to less than about 20\%. This confirms that our results are
 not too sensitive to the uncertainties in the luminosity and redshift
 distribution of Compton thick AGN.

The integrated SMBH mass density as a function of redshift 
can then be calculated starting from
a local value for $\rho_{\rm BH,0}$ (see section~\ref{sec:local}):
\begin{equation}
\label{eq:rhobh_z}
\frac{\rho_{\rm BH}(z)}{\rho_{\rm BH,0}}=1-\int_0^{z}\frac{\Psi_{\rm BHAR}(z')}{\rho_{\rm BH,0}}\frac{dt}{dz'}dz'.
\end{equation}
In this way, 
for any given $\phi(L_{\rm X},z)$ and bolometric correction  (thus
$\Psi_{\rm BHAR}(z)$), the exact
shape of $\rho_{\rm BH}(z)$ then depends only on
two numbers: the local black holes mass density $\rho_{\rm BH,0}$ and
the (average) radiative efficiency $\epsilon_{\rm rad}$. This fact has
an interesting, and somewhat surprising, consequence: even before
making any assumption about the distribution of the Eddington ratio in
the AGN population, we are able to constrain the (mass weighted)
average radiative efficiency, in the following way. A firm {\it lower}
limit to  $\epsilon_{\rm rad}$ can be established using the Soltan
argument. Specifically, it is enough to look for the value of the
radiative efficiency that makes the integral
$\int_0^{z}\frac{\Psi_{\rm BHAR}(z')}{\rho_{\rm BH,0}}\frac{dt}{dz'}dz'$
of eq.~(\ref{eq:rhobh_z}) larger than unity. For our choice of the
local SMBH mass function and AGN bolometric luminosity function this
lower limit is approximately equal to 0.065.

However, when starting from the local mass density to compute its
evolution, the {\it shape} of $\rho_{\rm BH}(z)$ turns out to be very
 sensitive to
the average radiative efficiency. The very fact that, in much the same
way as the SFR density, also the black hole accretion rate density,
$\Psi_{\rm BHAR}(z)$ has
a well pronounced maximum (around $z\sim 2$), implies that, if the
adopted value of $\epsilon_{\rm rad}$ were too high, the black hole mass
density would not decrease any longer towards higher redshift. This
is clearly shown in the upper two panels of
Figure~\ref{fig:zev_100_083}, where we show the evolution of the total
SMBH mass density computed for two different values of the accretion
efficiency (and, correspondingly, of the radiative efficiency):
$\eta=0.1$ (corresponding to $\epsilon_{\rm rad}\simeq 0.086$) and
$\eta=0.083$ ($\epsilon_{\rm rad}\simeq 0.07$).

The same issue can be considered from a different point of view. Let us now
 define a global, mass weighted average Eddington ratio for the
entire SMBH population as:
\begin{equation}
\label{eq:aveledd}
\Lambda(z) \equiv 4.38 \times 10^{8}
\epsilon_{\rm rad} \left[\frac{\Psi_{\rm BHAR}(z)}{\rho_{\rm BH}(z)}\right]
\end{equation}
The redshift evolution of $\Lambda(z)$ is shown in the two bottom
panels of  Fig.~\ref{fig:zev_100_083}\footnote{Similar calculations were shown
already in Wang et al. (2008), where $\Lambda$ was interpreted as the
product of an ``active'' BH duty cycle times its average Eddington
rate $\langle\dot m \rangle$; 
with our method, that calculates the accretion rate distribution
directly, there is no need to introduce a fixed $\langle\dot m
\rangle$ for
active black holes. As we will discuss in \S~\ref{sec:active}, the
very concept of active SMBH becomes purely observation-dependent. For
a discussion of how to interpret lifetimes and duty-cycles see also M04.}. There
it is apparent that,
while for the low radiative efficiency case (lower right panel) 
the average Eddington ratio is allowed to
increase up to the highest redshift for which reliable X-ray
luminosity data are available
($z\sim 4$), increasing the efficiencies by less than 20\% modifies the
evolution of the $\Lambda$ substantially (see bottom left panel), 
being it almost flat, and possibly decreasing at $z\ga
3$, an obvious consequence of the flattening of the corresponding mass 
density (top left panel). 
The physical explanation for this is straightforward: if we
calculate the evolution of the SMBH mass density with a too high
radiative efficiency, we fail to account for the mass locked up
in SMBH in the local Universe, and we are forced to explain a
``primordial'' population of very weakly active (or even dormant)
black holes, of substantial density in place by $z\sim
4$ (in the example of
Fig.~\ref{fig:zev_100_083}, left panel, amounting to about 10$^5$
$M_{\odot}$ Mpc$^{-3}$). This puts
already some strain on existing models for the black holes seed population
(see e.g. Trenti \& Stiavelli 2006; Volonteri, Lodato and Natarayan
2007). Models
in which the primordial black holes originate from stellar mass
progenitors (remnants of the first, PopIII, stars;
Abel, Bryan \& Norman 2000; Bromm, Coppi \& Larson 2002), 
predict negligible BH mass density at high redshift (say, less
than 10$^4$ $M_{\odot}$ Mpc$^{-3}$ at $z>5$), and are incompatible with an
average, mass weighted radiative efficiency larger than about 0.075
(in very good agreement with the recent estimates of Shankar et al. 2008b),
or, correspondingly, with an accretion efficiency $\eta \ga 0.09$.

Such a tight {\it upper} limit on the radiative efficiency can be somewhat relaxed if:
 (a) the local SMBH mass density has been overestimated (see the
discussion in Merloni et al. 2004); (b) the Compton Thick population
has a very different redshift distribution 
from that assumed in the Gilli et al. (2007) XRB synthesis models
(with more heavily obscured high luminosity objects at high redshift);
(c) the primordial BH seeds are
indeed very massive, resulting from the direct collapse of
supermassive stars \cite{koushiappas:04,begelman:06,lodato:07}.
 
On the other hand, the strict {\it lower} limit on the efficiency could also
be relaxed if the local SMBH mass density had been underestimated, or
if there was a substantial population of ``wandering'' SMBH, i.e. objects
that have been kicked out of their galactic host, most likely 
for the recoil experienced 
due to asymmetric gravitational wave emission subsequent to a
binary black hole merger event, {\it after}  having accreted
substantial amount of matter. Those black holes would then be computed
in the AGN luminosity functions, but not among the local relic
population that contributes to $\rho_{\rm BH,0}$. A discussion of the
relevance of this effect on the mass function of SMBH can be found in
Volonteri (2007).

Summarizing, we can define: (i)  a normalized local SMBH mass density 
$\xi_0\equiv \rho_{\rm BH,0}/(4.3\times 10^5 M_{\odot} {\rm
  Mpc}^{-3})$; (ii) the normalized SMBH mass density at
$z=z_i$ (where $z_i$ is our current observational horizon, i.e. the
maximum redshift at which AGN luminosity functions are well known)
$\xi_i\equiv \rho_{\rm BH}(z=z_i)/\rho_{\rm BH,0}$ and (iii) the
normalized mass density of SMBH ejected from galactic nuclei due to the
gravitational wave recoil after mergers, $\xi_{\rm lost}\equiv
\rho_{\rm BH,lost}/\rho_{\rm BH,0}$. Then, our calculations show that
the average radiative efficiency is constrained to be
\begin{equation}
\label{eq:radeff_const}
\frac{0.065}{\xi_0(1+\xi_{\rm lost})}\la \epsilon_{\rm rad}
 \la \frac{0.070}{\xi_0(1-\xi_i+\xi_{\rm lost})}.
\end{equation}

\begin{figure}
\centering
\psfig{figure=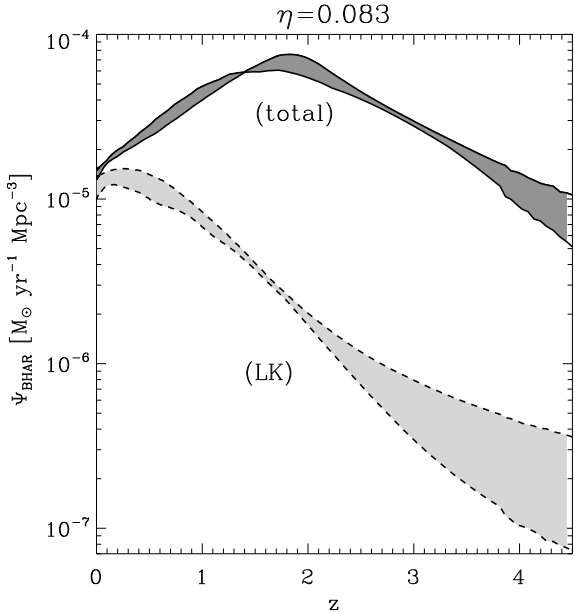,height=8.5cm}
\caption{Redshift evolution of the accretion rate density onto
  supermassive black holes
$\Psi_{\rm BHAR}(z)$. Darker shaded area (solid lines) shows the total
accretion rate, while the lighter ones (dashed lines) shows the mass
accreted in the LK mode only. This corresponds to about 20-27\% of
the total mass growth of the SMBH population. }
\label{fig:zev_mdot_083}
\end{figure}

Finally, we can calculate explicitly how much mass have SMBH 
 accumulated in different modes. Specifically, we find that
between about 73\% and 80\%  (depending on the choice of the
luminosity functions) of the relic black hole mass density has
been accumulated in radiatively efficient modes (either HR or HK), while
only 20\% to 27\% in the low radiatively efficient (LK) mode, even if
most of the time in the life of a SMBH is likely spent in this latter
mode \cite{cao:07,hopkins:06a}. The accretion rate density evolution
for both the entire population and for the SMBH in the LK mode is
shown in Figure~\ref{fig:zev_mdot_083}.

\begin{figure*}
\centering
\psfig{figure=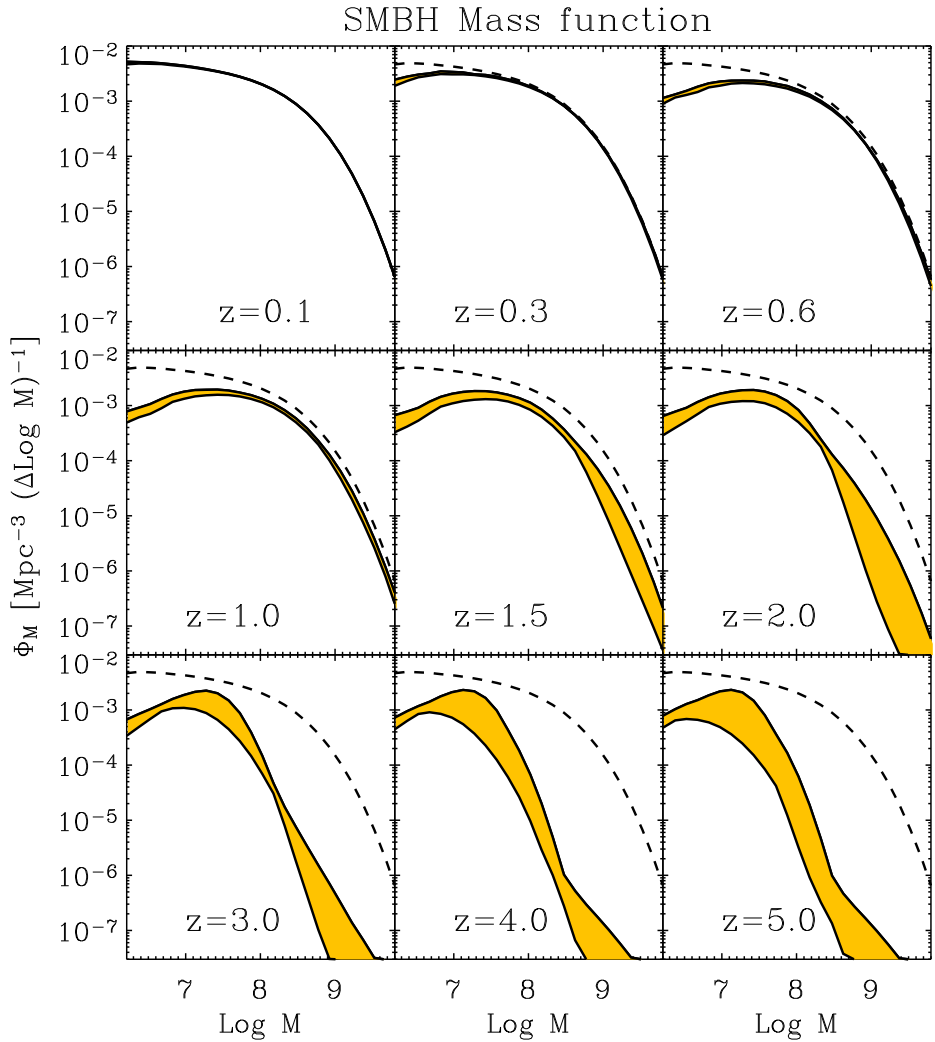,height=15cm}
\caption{Redshift evolution of the SMBH mass function. Shaded areas
  represent the uncertainties calculated based on the estimated
  uncertainties in the luminosity functions entering our
  calculations (see section~\ref{sec:mf} for details). 
The continuity equation for SMBH growth was integrated
assuming a constant accretion efficiency of $\eta=0.083$, which for the
adopted model of the various accretion modes (\S~\ref{sec:modes})
corresponds to a average radiative efficiency of $\sim 0.07$. In all
panels the $z=0.1$ mass function is shown as a dashed line for
reference. Black hole masses are measured in units of solar masses.}
\label{fig:mf_083}
\end{figure*}

\subsection{The SMBH mass function evolution}
\label{sec:mf_sub}
So far, we have only discussed integral constraints, i.e. those
obtained from the study of the evolution of the integral of the mass
(or luminosity) function. However, the solution of
eq.~(\ref{eq:continuity}) provides us with the full evolution of the
SMBH mass function, and we proceed now to discuss it in detail.

The SMBH mass function evolution is shown in Figure~\ref{fig:mf_083} for our
fiducial case of $\epsilon_{\rm rad}=0.07$. In the different panels, each
corresponding to a different redshift, the dashed lines show the low redshift
($z=0.1$) mass function as a reference, while the shaded areas
represent the calculated mass functions with their own uncertainties.   

The so-called ``downsizing'' is apparent in the first four or five
panels (i.e. up to redshift $\sim 1.5-2$): we clearly see how in this
time interval the mass function grows much more substantially at the low-,
rather than at the high-mass end. This is to be expected, in the light
of the results of Heckman et al. (2004). There,
on the basis of a sample of about 23,000 SDSS type II AGN 
for which both an estimate of the black hole mass (via a
measurement of the velocity dispersion of the host) and of the
intrinsic accretion rate (via measures of the narrow line luminosity)
were available, it was shown that it
is only the small mass black holes that are growing rapidly
at $z\sim 0.2$. Similar
trends, indicating higher Eddington ratios (and smaller growth times,
see below) have been also found recently in the SDSS quasar sample
\cite{netzer:07}, dominated by un-obscured, luminous AGN at higher
redshift. We stress, however, that our results, based on a much more
statistically complete census of the AGN population (thanks to the
hard X-ray selection), is much less
prone than previous ones to selection biases.

Differently from what previously found in M04, we note here that by 
 the epoch of the peak of the AGN/QSO activity ($z\sim 2$), 
the high-end of the mass function decreases significantly, too, in
coincidence with the observed drop of the number density of bright QSOs at those
redshifts \cite{fan:01,wall:05,hasinger:05,richards:06,silverman:08,brusa:08}.

\begin{figure}
\psfig{figure=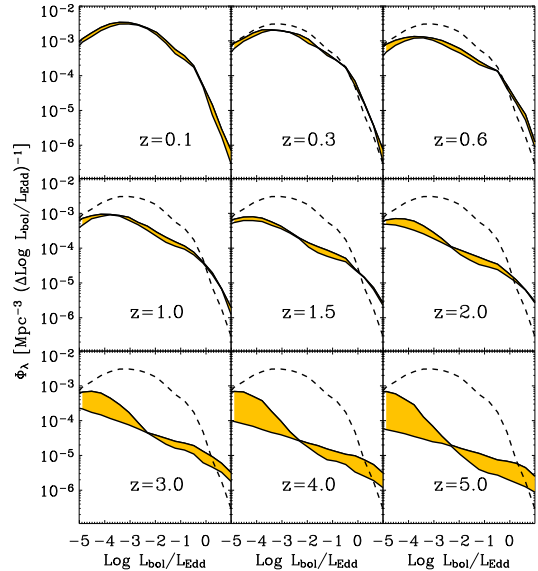,height=8.5cm}
\caption{Differential distribution of the Eddington ratio $\lambda=L_{\rm
    bol}/L_{\rm Edd}$ at different redshifts. In all panels,
  the distribution is shown as shaded areas, while dashed lines
  represent the $z=0.1$ case, shown as a reference.}
\label{fig:mdotf_083}
\end{figure}


\subsection{The accretion rate distribution function: downsizing and its
  reversal }
\label{sec:mdotf}
As we have stressed a few times already, our calculations do not
assume any particular distribution of accretion rates for AGN, rather derive
it consistently from the observed mass and luminosity functions (see
\S~\ref{sec:methods}). It is thus interesting to show the evolution of
the (Eddington-scaled) accretion rate function, here defined as:
\begin{equation}
\Phi_{\lambda}=\int \Phi_{M, \dot M}[M,\dot M(M,\lambda)]\frac{\partial
  \dot M}{\partial \lambda} d\log M,
\end{equation}
where $\Phi_{M, \dot M}(M,\dot M)$ is given by
eq.~(\ref{eq:joint_mdot}), the relationship between $\dot M$ and
$\lambda$ is simply given by $\lambda=\epsilon_{\rm rad}\dot M
c^2/L_{\rm Edd}$, and $\lambda$ is related to the accretion rate via
eq.~(\ref{eq:radeff}). We show the Eddington ratio distribution function
evolution in Figure~\ref{fig:mdotf_083}. The various panels make
evident the change in the shape of the distribution, which results
from  the combined evolution of the X-ray luminosity function (a form of
luminosity dependent density evolution is noticeable in the evolution
of $\Phi_{\lambda}$ as well) and of the mass function itself. Overall,
the trend is for a progressive flattening of the accretion rate
distribution function with increasing redshift, i.e. of a an
increasing relative importance of highly accreting objects, as it is
reasonable to expect on physical grounds if the SMBH population grows
from small mass high-redshift seeds.

\begin{figure*}
\centering
\psfig{figure=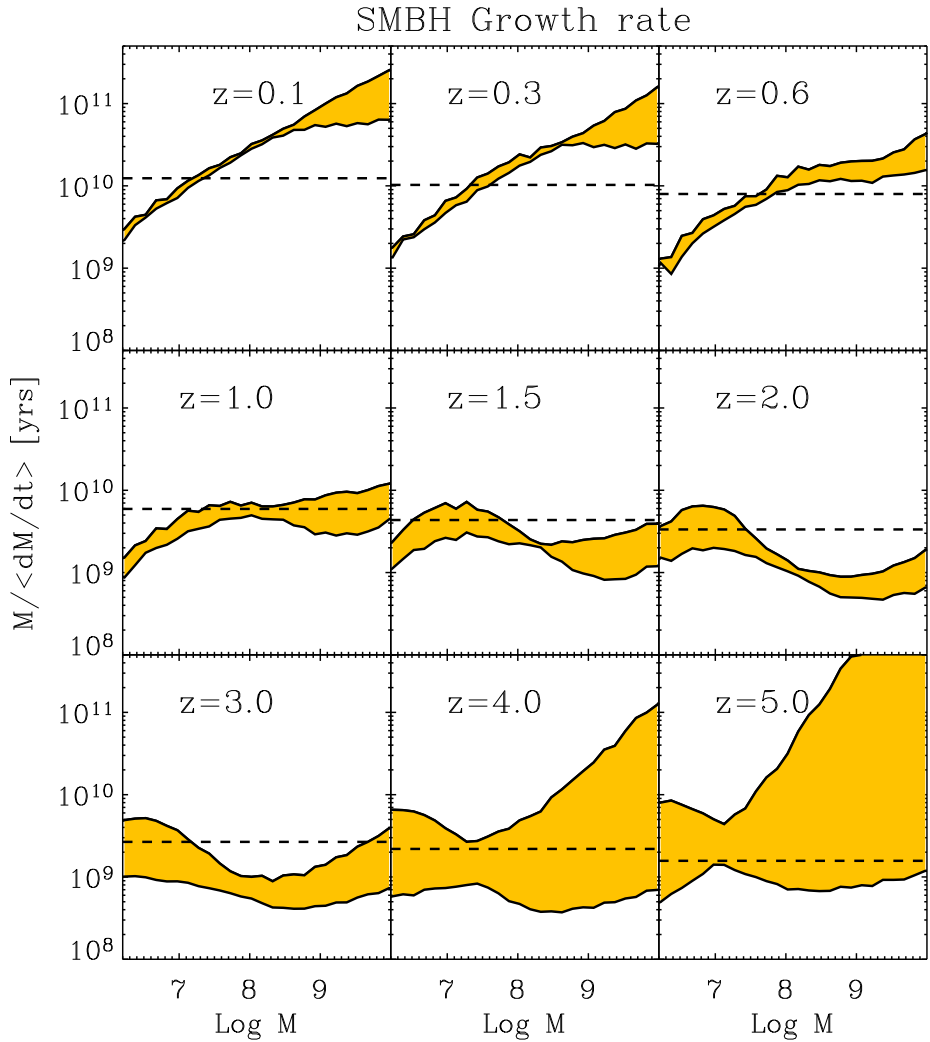,height=15cm}
\caption{Redshift evolution of the growth time (in years) as a function of
black hole mass. In each panel, a horizontal dashed line marks the age
of the Universe at that particular redshift. Black hole masses are
measured in units of solar masses.}
\label{fig:mass_frac_083}
\end{figure*}

The evolution of the Eddington ratio distribution function shown in
Figure~\ref{fig:mdotf_083} does not contain any information about the
typical accretion rate as a function of black hole mass. A useful way
to present this information is that of showing the specific {\it growth
rate} as a function of SMBH mass. We have computed this quantity by
taking the ratio of the black hole mass to the average accretion rate
$\langle \dot M(M)\rangle$. Such  a ratio $M/\langle \dot M\rangle$
defines a timescale, the so-called {\it growth time}, or mass doubling
time, as it measures the time it would take a black hole of mass $M$
to double its mass if accreting at the currently measured rate of
$\dot M$. The redshift evolution of the growth time (in years) as a function of
black hole mass is shown in Figure~\ref{fig:mass_frac_083}. As a
reference, we show in each panel the age of the Universe as a
horizontal dashed line. Black holes with growth times longer than the
age of the Universe are not experiencing a major growth phase, which
must have necessarily happened in the past. On the contrary, objects
with growth times shorter than that are actively growing. 

The $z=0.1$ growth time distribution is in good agreement, both in
shape and normalization, with that first measured in the above
mentioned work of Heckman et
al. (2004). The positive slope of the average growth time vs. $M$
is one of the most direct evidence of a SMBH ``downsizing'', i.e. of the fact
that in the local universe only small mass black holes are actively growing. 
Interestingly enough, a similar trend has been since unveiled for star
forming galaxies \cite{feulner:05,noeske:07,perez:08}, where the ratio of galaxy mass
$M_{\rm gal}$ to star formation rate is studied as a function of $M_{\rm gal}$.
It is clear that a direct comparison of the SMBH growth times so
measured with those of the stars in galaxies of various
morphologies/ages/masses may hold the key for a thorough understanding
of the SMBH-galaxies co-evolution, and we aim at presenting such a
comparison in a following paper.

The redshift evolution of the growth times distribution can be used to
identify the epochs when black holes of different masses grow the
larger fraction of their mass. Figure~\ref{fig:mass_frac_083} shows
that as we approach the peak of the black hole accretion rate density
($z\sim 1.5-2$, see fig.~\ref{fig:zev_mdot_datasfr}), we witness the
rapid growth of high mass black holes, too. The growth time
distribution as a function of mass becomes flat there, and sinks
below the corresponding age of the Universe. This is the period of
rapid, widespread growth of the entire SMBH
population. Correspondingly, above this
redshift, the ``downsizing'' trend seems to disappear.

\begin{figure*}
\centering
\begin{tabular}{ll}
\psfig{figure=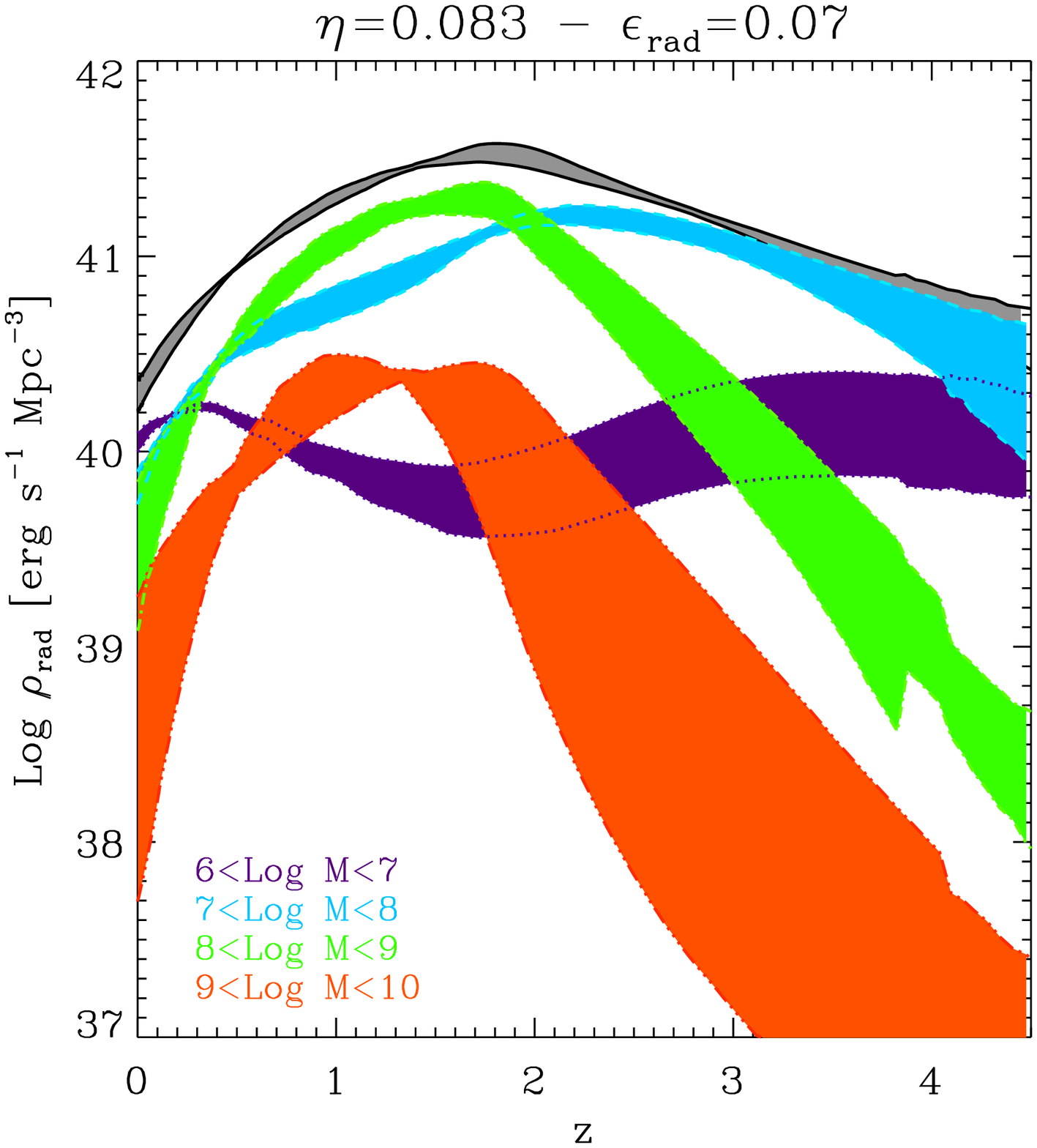,height=8.5cm}&
\psfig{figure=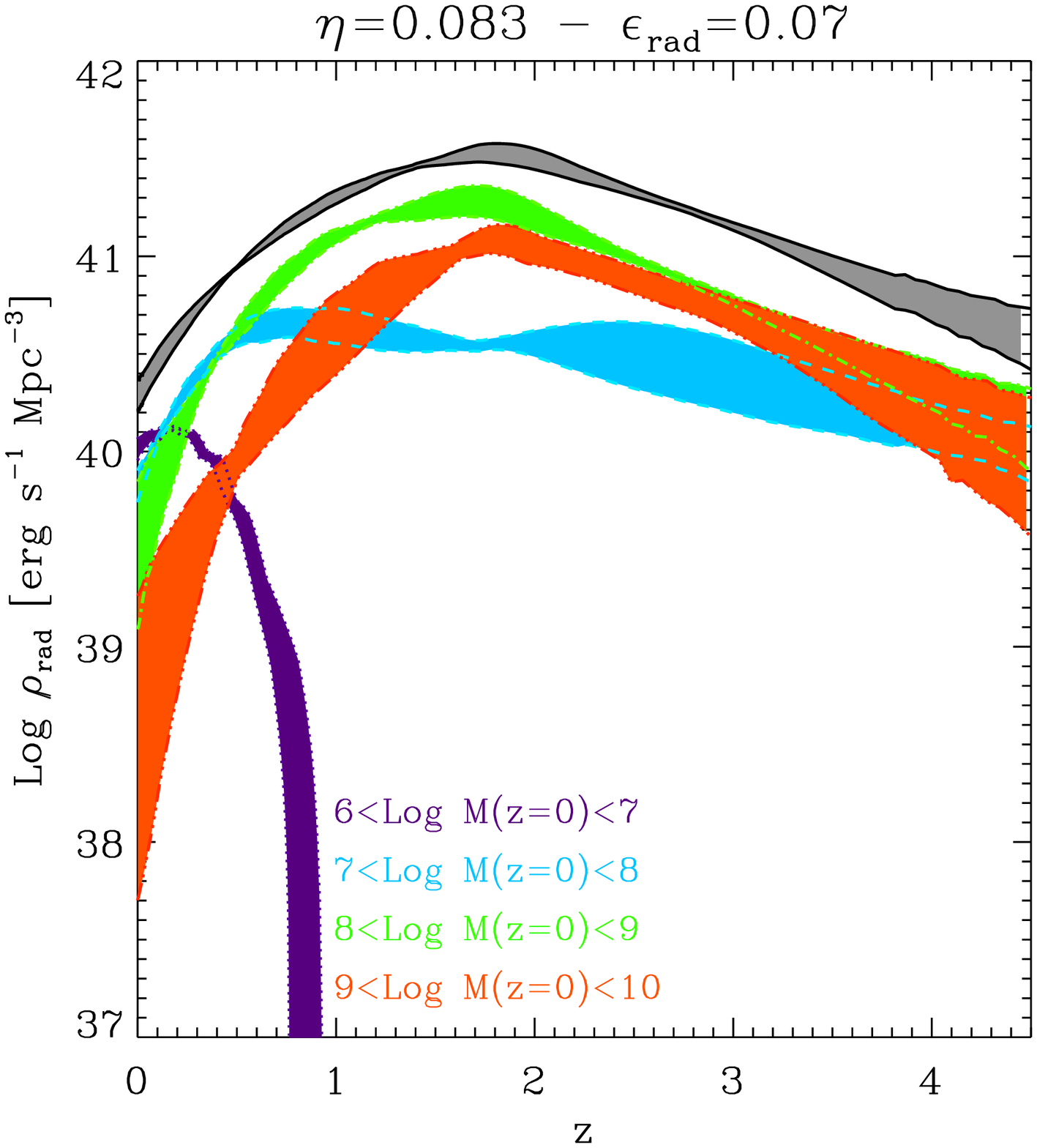,height=8.5cm}\\
\end{tabular}
\caption{Redshift evolution of the radiative (bolometric) energy
  density of AGN. Gray shaded area (between solid lines) 
  is the total integrated energy density,
  while orange (triple-dot-dashed), green (dot-dashed), cyan (dashed)
  and purple (dotted) areas are the radiative energy density emitted by black
holes with $\log M>9$, $9>\log M>8$, $8>\log M>7$ and $7>\log M>6$,
respectively (all in units of solar masses). The right hand plot shows
the same quantities for objects divided according 
to their instantaneous mass, while the right hand one is for objected
partitioned according to their final ($z=0$) mass.}
\label{fig:zevrad_083}
\end{figure*}

Finally, we show in Figure~\ref{fig:zevrad_083} the redshift evolution of
the radiated (bolometric) energy density split into various black
hole mass bins. 
On the left, we divide objects according to their mass
at each redshift. The low redshift downsizing trend is here apparent
in the rapid decline of the radiated energy density of the most
massive black holes ($\log M>9$), which becomes less dramatic when
less massive holes are considered. While in the local Universe
(i.e. for $z\la0.3$) the
larger contributors to the total 
radiative energy density are the numerous, small, rapidly accreting SMBH (with
$\log M<7$), above redshift one most of the radiative energy is
produced by larger black holes, typically with masses between $10^8$
and $10^9$ solar masses. As we move to higher redshift, the numbers of
black holes with $\log M>8$ declines, and the radiative energy density
is dominated again by smaller black holes with $8>\log M>7$. The most
massive black holes, those with $\log M>9$, never dominate the
radiative energy density injection, essentially because of their small
numbers, being always in the exponentially declining part of the
mass function, well above its knee (see Figure~\ref{fig:mf_083}).

As we discussed in the Introduction, cosmological studies of SMBH
evolution offer the advantage, with respect to those of galaxies, that
SMBH do not undergo any morphological transformation as they grow. 
Thus, once 
enough information is gathered to solve the continuity
equation~(\ref{eq:continuity}), we can trace back the evolution of any
black hole (or of any population of black holes of any given mass at a
particular redshift). 
Indeed, it was by comparing the growth histories of black
holes of different masses today that Marconi et al. (2004) first
noticed what they called the ``anti-hierarchical'' evolution of the SMBH
population (see also M04; Shankar et al. 2008b). Obviously, we
can do the same here. On the right hand side of
Figure~\ref{fig:zevrad_083}, we show the redshift evolution of the
radiative (bolometric) energy density split into bins of different
black hole masses {\it today}, i.e. at $z=0$. 

From this we see that the progenitors of local SMBH with masses in the
range $8< \log M<9$ dominate the AGN radiative energy output between
$z\sim 0.3$ and $z\sim 3$, when the progenitors of the largest black
holes today (those with $\log M>9$) started to become at least equally
important.

\subsection{Selection effects: mass function of ``active'' black holes}
\label{sec:active}
From our discussion so far, it should be clear that a complete picture
of the growth and evolution of the SMBH population cannot 
be achieved without appreciating the full extent of the wide
distribution of both masses and accretion rates, a point often
overlooked in either semi-analytic or numerical models.
As a consequence, the very concept of a clear distinction between
``active'' and ``inactive'' black holes can be misleading, once we
consider within the same scheme black holes accreting close to the
Eddington limit and those with an Eddington-scaled accretion rate as
small as, say, $10^{-4}$. 

\begin{figure}
\centering
\psfig{figure=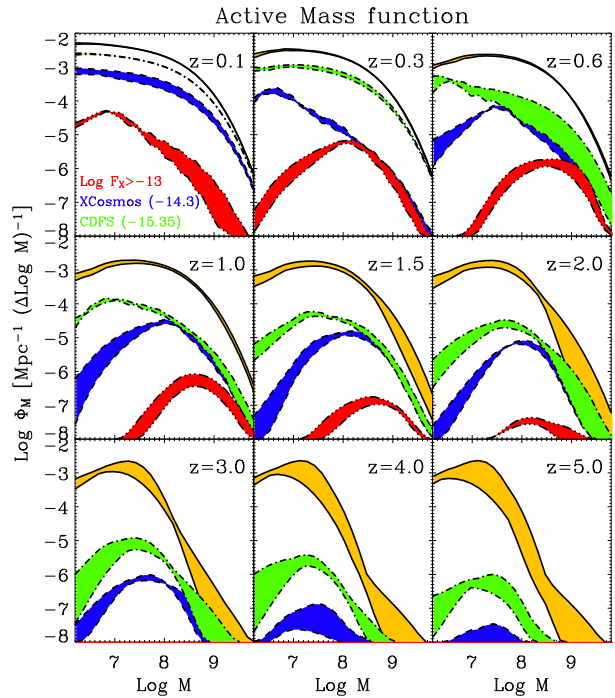,height=10cm}
\caption{Mass functions of active SMBH with observed 2-10 keV flux
  above  $10^{-13}$ (red), $10^{-14.3}$ (blue) and $10^{-15.35}$
(green) erg/s/cm$^2$. As a reference, the total SMBH mass function is
also shown as a yellow area between solid lines, from
fig.~\ref{fig:mf_083}.}
\label{fig:mflux_083}
\end{figure}

Thus, we advocate here more operational definitions of ``active'' black
hole, the easiest and most straightforward of which are those based on
nuclear flux limits. As an illustrative example, we plot in
Figure~\ref{fig:mflux_083} the mass functions of SMBH whose {\it
  observed} X-ray flux exceeds some fixed limits. In particular,
alongside the total mass function, reproduced here from
Fig.~\ref{fig:mf_083} as a term of comparison, we plot the mass
functions of all SMBH with an observed (i.e. absorbed) 2-10 keV  flux
above $10^{-13}$ (red), $10^{-14.3}$ (blue) and $10^{-15.35}$
(green) erg/s/cm$^2$. The latter two limits have been chosen to match
the observational limits of recent deep (Chandra Deep Field South,
CDFS; Giacconi et al. 2002) and medium-deep 
(XMM-COSMOS, XCosmos; Hasinger et al. ) surveys. Due to the
relative steepness of the accretion rate functions at low redshifts,
the mass functions of active black holes so defined are extremely
sensitive to the flux limit defining what ``active'' means. 

This fact implies that, when trying to determine the SMBH mass function from
a survey that selects AGN above a given flux limit, one should
carefully account for the bias introduced by the fact that SMBH of any
mass always have a broad distribution of accretion rates
\cite{babic:07}. 
For obvious reasons of simplicity and coherence, we discuss here hard X-ray
flux limits, as we use the hard X-ray luminosity function to unveil the
SMBH evolution, and the X-ray background to constrain the relative
numbers of obscured and unobscured sources. However, it is clear that,
provided robust bolometric corrections are available, similar
calculations can be converted into flux limits in any band. For
example, Greene and Ho (2007) have recently presented the mass
function of active black holes, identified as broad H$\alpha$ line
emitters (in the redshift range $0.1\la z\la 0.3$) in the SDSS
database. Comparison with the soft X-ray selected AGN luminosity
function of Hasinger et al. (2005) showed that their sample approach
completeness at luminosity corresponding to a (rough) X-ray flux limit
of 10$^{-13}$ erg/s/cm$^2$. Indeed, the mass function we show at low
$z$ in Fig.~\ref{fig:mflux_083} for such a flux limit is in quite good
agreement with that reported in Greene and Ho (2007), once the fact
that our active mass functions do include both obscured and unobscured
objects is accounted for. 

\begin{figure}
\centering
\psfig{figure=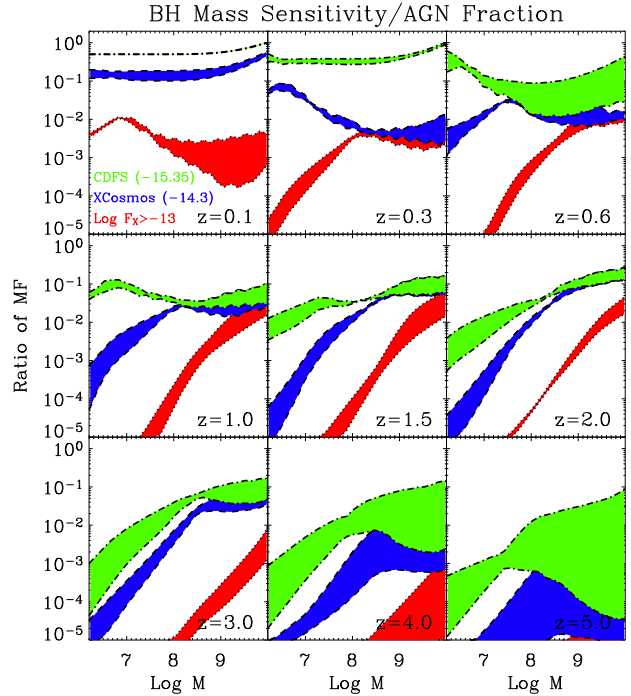,height=10cm}
\caption{Ratio of SMBH mass function above a given flux limit to the
  total mass function. The various shaded areas correspond to
  different observed 2-10 keV flux limits, the same as those used in
  figure~\ref{fig:mflux_083}, i.e. 
  $10^{-13}$ (red), $10^{-14.3}$ (blue) and $10^{-15.35}$
(green) erg/s/cm$^2$.}
\label{fig:mflux_ratio_083}
\end{figure}

In Figure~\ref{fig:mflux_ratio_083} we show the ratio of the
various ``active'' mass functions to the total SMBH mass
function. These curves can be simply interpreted as ``selection
functions'' for the black hole mass of AGN surveys of corresponding
hard X-rays flux limits. However, as far as we can claim that, at all
redshifts considered here, there is a one-to-one correspondence
between SMBH and galaxies mass (as it was implicitly assumed to derive the
$z=0$ mass function), then the various curves of
Figure~\ref{fig:mflux_ratio_083} could be also interpreted as ``AGN
fractions'' as a function of SMBH (and galaxy) mass. Once again, due
to the main tenet of our method of calculation, namely the fact that
growing supermassive black holes display a very broad accretion rate
distribution at any given mass,  the very concept of ``AGN fraction''
is strongly dependent on the sensitivity limit that defines
what an AGN is. In general, the AGN fraction tends to be highest at
the high mass end, and declines at low masses. A deep X-ray survey
like the CDFS has a quite flat 'mass sensitivity' (or mass fraction)
at $z\sim 1$, while the shallower XMM-COSMOS at the same redshift
starts declining already for $\log M<8$. At higher redshift even the
deepest surveys are highly incomplete (in a SMBH mass sense), and the
incompleteness rises steeply with decreasing mass.
In the case of spectroscopic studies (see e.g. Decarli
et al. 2007), where the
definition of an active galactic nucleus relies on the capability of
spatially and spectrally resolve AGN-excited emission lines, the AGN
fraction will be a more complicated function of SMBH (and galaxy) mass
than shown in Figure~\ref{fig:mflux_ratio_083}.

\section{The evolving kinetic luminosity function of AGN}
\label{sec:kinetic}
So far, we have focused our attention on the growth of the SMBH
population under the assumption that, whatever the mode of accretion
of a black hole, it s possible to establish a
unique correspondence between radiated (bolometric) luminosity and
rate of change of black hole mass (see section~\ref{sec:modes}). A
choice of local black hole mass function, an expression for the
bolometric correction, and a choice of a value for the accretion
efficiency was all that was needed to extract the useful information
on the SMBH growth from the observed X-ray luminosity function.

However, the growth of supermassive black holes through mass
accretion is accompanied by the release of enormous amounts of
energy which, if it is not advected directly into the hole (see
e.g. Narayan \& Yi 1995), can not only be radiated away, 
but also disposed of in kinetic form through
powerful, collimated outflows or jets, as observed, for example, in Radio
Galaxies. To reiterate the point made before (\S~\ref{sec:modes}), it
is only by assessing the relative importance of the radiative and
kinetic energy output in either ``low kinetic'' (LK), ``high
radiative'' (HR) and ``high kinetic'' (HK) modes that a full picture
of the effects of SMBH growth on the environment will possibly emerge.

This is important, as we expect that radiative and kinetic feedback
differ not only on physical grounds,  
in terms of coupling efficiency with the
ambient gas, but also on evolutionary grounds. The anti-hierarchical
growth of the SMBH population discussed in sections~\ref{sec:mf_sub}-\ref{sec:active}
necessarily implies a delay between the epoch of
Quasar (HR) dominance and that of more sedate LK mode, 
such that kinetic energy feedback plays an
increasingly important role in the epoch of cluster formation and
virialization. 

In this section, we present a method to complement the information on
the cosmological growth of the SMBH population described in the
previous sections with an analysis of the evolving kinetic luminosity
function of AGN. As in Merloni (2004), we will make use of the radio
luminosity function of AGN as the main observable for our
study. However, differently from the choice made there, we will focus
our attention on the {\it intrinsic radio core} luminosity function. The main
reason for this choice is twofold. 

On the one hand, the observed
relation between BH mass, radio and X-ray luminosity (the fundamental
plane of active black holes, MHD03) that defines the physical state of LK mode
objects is based on the observed (5 GHz) radio core emission, not on
the extended one. Also,
Merloni and Heinz (2007) have shown that the intrinsic radio core
luminosity of an AGN jet is a good indicator of its total kinetic
power, once the effects of relativistic beaming are taken into
account. On the other hand, the physical model for the radiative emission in
the core of a relativistic jet is relatively well established
\cite{blandford:79,heinz:03}, 
as opposed to the complicated physics of
radio emission and particle acceleration in the large scale lobes of a
radio jet, where environmental effects do necessarily play an important
role. Obviously, 
the price to pay for the use of radio core emission as a tracer of the
jet power, is that relativistic beaming may severely affect the derived
parameters, and must be taken into account.

\begin{figure*}
\centering
\begin{tabular}{ll}
\psfig{figure=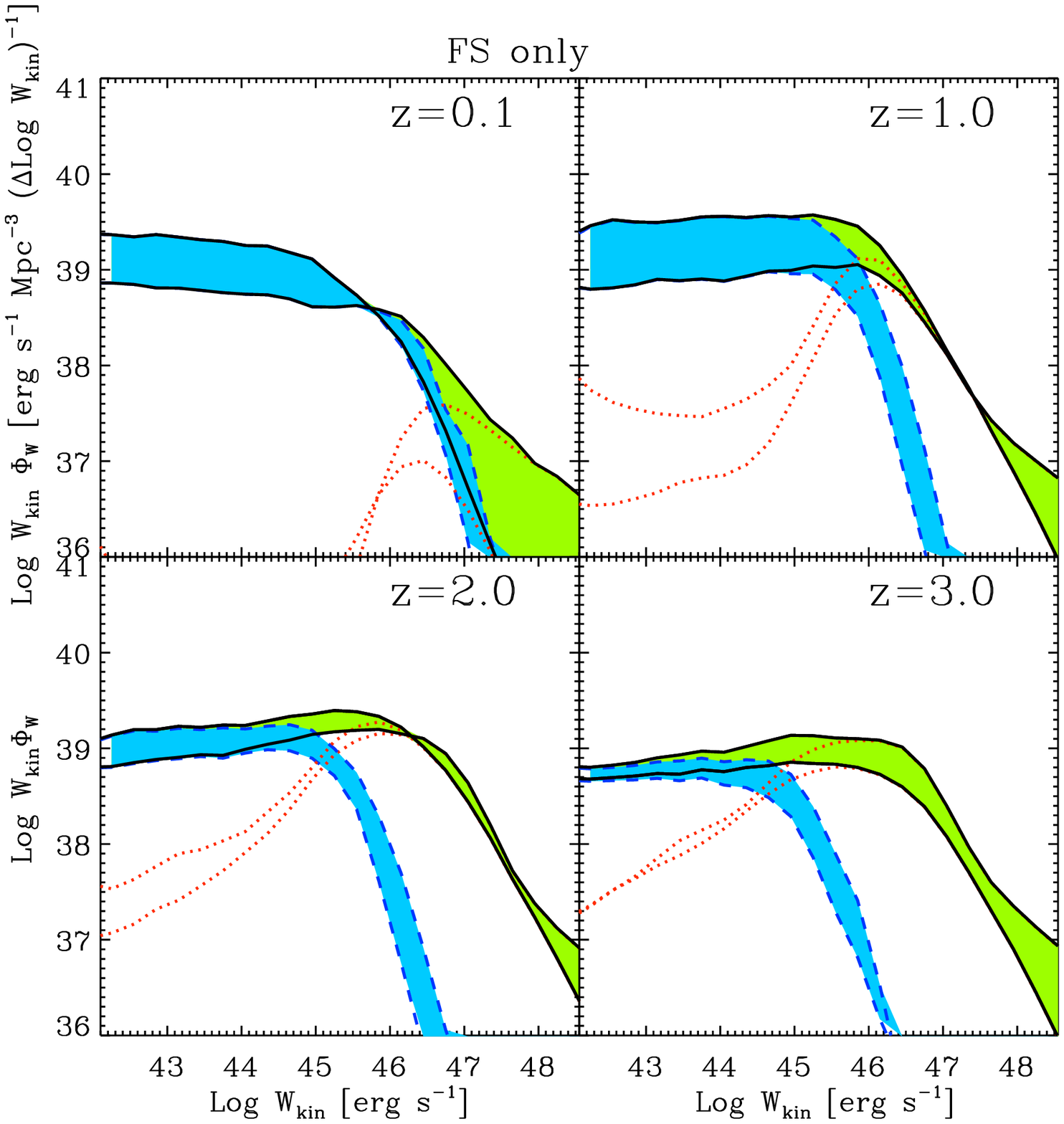,height=8.5cm}&
\psfig{figure=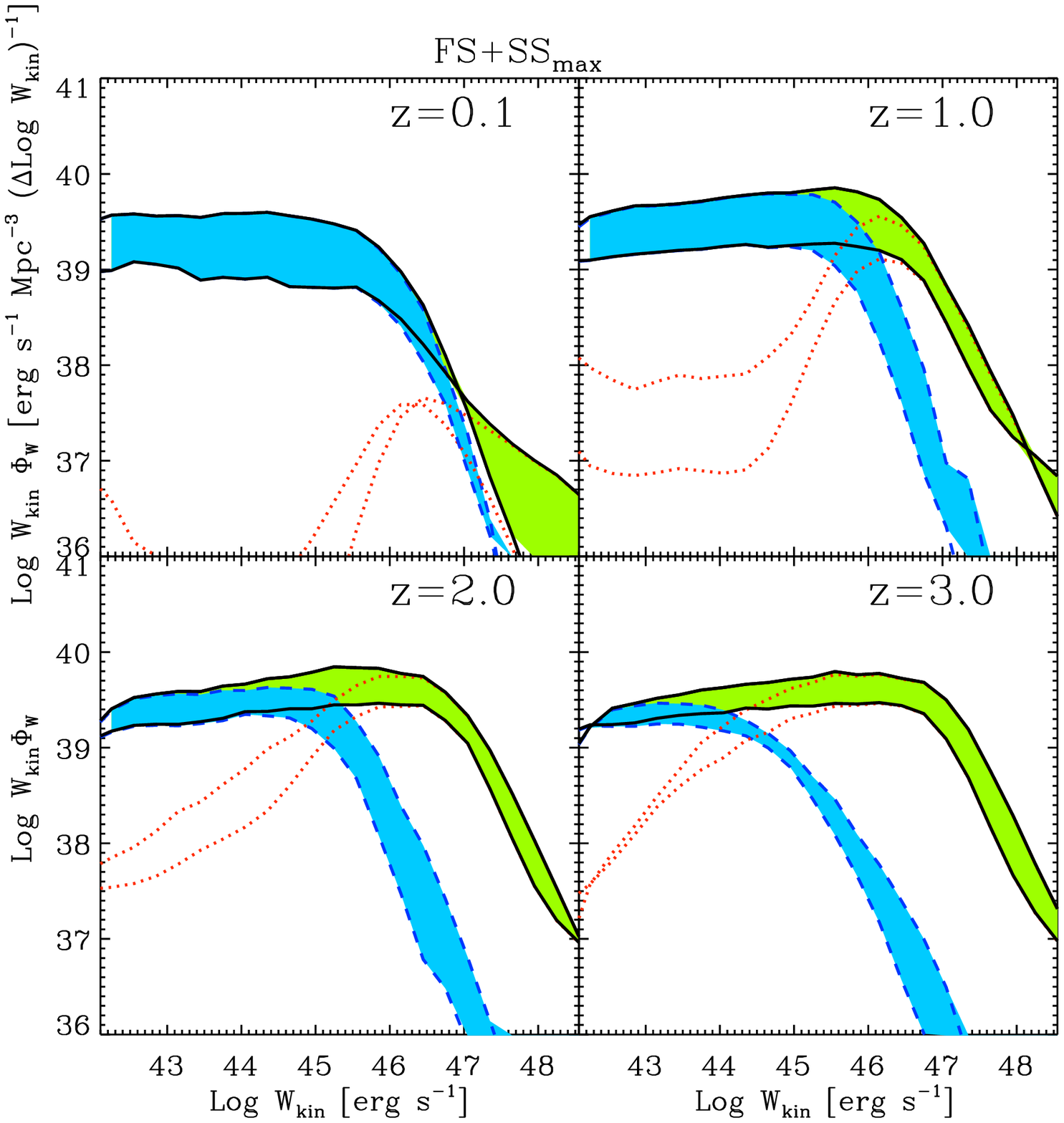,height=8.5cm}\\
\end{tabular}
\caption{Kinetic luminosity function of AGN jets (here plotted as
  $W_{\rm kin} \times \phi_{W}$) at four different redshifts. In each
  panel the green area between the black solid lines represents
the total kinetic luminosity function, with the uncertainty reflecting
the difference in the luminosity functions between Dunlop \& Peacock
(1990) and De Zotti et al. (2005). The blue area between blue dashed
lines represents the contribution from sources in the LK mode, while
the dotted lines mark the contribution from sources in the HK
state. In the {\it Left plot} (FS) the jet core luminosity function
includes only flat spectrum sources, while in the {\it right plot}
(FS+SS$_{\rm max}$) a maximal contribution from hidden cores in steep
spectrum sources is included (see text for details).}
\label{fig:lkinf}
\end{figure*}

 In the classical model by Blandford \& Koenigl (1979), the
flat spectrum radio synchrotron emission of a compact jet core is
produced by a superposition of self-absorbed synchrotron spectra, each
from a different region in the jet. The model predicts a dependence of
the monochromatic 
radio luminosity $L_{\nu}$ on jet kinetic power $W_{\rm kin}$ of the form
$L_{\nu} \propto W_{\rm kin}^{17/12}$.  More generally,
Heinz \& Sunyaev (2003) showed that {\em any} scale invariant jet model
producing a power-law synchrotron spectrum with index $\alpha_{\nu}$
{\em must} obey the relation $L_{\nu} \propto W_{\rm
kin}^{(17+8\alpha_{\nu})/{12}}M^{-\alpha_{\nu}}$. By studying a sample
of nearby low-luminosity radio AGN for which the kinetic power had been
measured, 
Merloni \& Heinz (2007) have shown that, once beaming is statistically
taken into
account, a tight relation between the 5 GHz intrinsic
radio core luminosity ($L_{\rm R}$) and the jet kinetic power holds, in the form: 
\begin{equation}
  W_{\rm jet}=W_0\left(\frac{L_{\rm
        R}}{L_0}\right)^{0.81}\;\; {\rm erg}\;{\rm s}^{-1}
  \label{eq:radiopower}
\end{equation}
where $W_0\simeq 1.6\times 10^{36}$ and $L_0=10^{30}$ erg s$^{-1}$,
and the intrinsic scatter is about 0.35 dex.
In \S\ref{sec:fslf}, we will use this relation to
construct the kinetic jet luminosity function from the observed flat
spectrum radio luminosity function $\Phi_{L}(L_{\nu})$ (abbreviated as
FSLF below). Similar, but complementary studies of the AGN kinetic
luminosity function evolution based on steep spectrum sources has been
recently carried out \cite{koerding:08,shankar:08a}. The results of both
these works are in reasonable agreement with those we present here.

\subsection{Beaming effects and the intrinsic luminosity function of
  radio jet cores}
\label{sec:fslf}
As pointed out in Heinz, Merloni \& Schwab (2007), we can use
eq.~(\ref{eq:radiopower}) and the observed FSLF to derive the
underlying kinetic luminosity function $\Phi_{W}$ of radio 
jet cores:
\begin{eqnarray}
  \Phi_{W}(W) &=& \phi_{\rm R}\left(L_{\rm R}(W)\right)\frac{dL_{\rm R}}{dW}
  \nonumber \\ &=& \phi_{\rm R}\left({L_0(\frac{W}{W_0})^{1/0.81}}\right)
  \frac{1}{0.81}\frac{L_0}{W_0}\left(\frac{W}{W_0}\right)^{\frac{1-0.81}{0.81}}
  \label{eq:kineticlumi}
\end{eqnarray}
Without loss of generality, and consistent with most observational
studies, we will use a broken power-law to describe $\phi_{\rm R}(L_{\rm R})$:
\begin{equation}
  \phi_{\rm R}(L_{\rm R})=\frac{\rho_0(z)}{L_{\rm
  R,c}(z)}\left[\left(\frac{L_{\rm R}}{L_{\rm R,c}(z)}\right)^{a_1} +
  \left(\frac{L_{\rm R}}{L_{\rm R,c}(z)}\right)^{a_2}\right]^{-1}
  \label{eq:lumi}
\end{equation}
In the above equation (\ref{eq:radiopower}), however, it is the {\it
  intrinsic} radio core luminosity that enters the determination of the
kinetic power. Since jets are relativistic, the observed
luminosity functions, $\phi'$ are always affected by Doppler
boosting.
Assuming all object have the same bulk Lorentz factor
$\gamma=(1-\beta^2)^{-1/2}$, the observed luminosity $L$ of a
relativistic jet is related to the emitted one, ${\cal L}$, via the
relation $L=\delta^p {\cal L}$, where $\delta$, the kinematic Doppler
factor of the jet is defined by $\delta=[\gamma(1-\beta cos
\theta)]^{-1}$. The exponent $p$ depends on the assumption about the
jet structure, and is here taken to be $p=2$ as appropriate for
continuous jets, rather than discrete ejections (for which $p=3$, see
Urry \& Padovani 1995).
If the radio sources have all a pair of oppositely directed jets, and
the jets are randomly oriented in the sky, 
the cosine of the angle between the velocity vector and the line of
sight, $\cos\theta$ have a uniform probability distribution, and the
observed sources will be those lying at angles between $\theta_{\rm
  min}$ and $\theta_{\rm max}$, corresponding to Doppler factors of
$\delta_{\rm min}$ and $\delta_{\rm max}$, respectively (see Urry and
Shafer 1984). We assume here that $\theta_{\rm max}=0^\circ$ and
$\delta_{\rm min}=1/\gamma$ (i.e. neglecting all strongly de-beamed
objects). 

Urry and Padovani (1991) have shown how, under the above assumptions,
an intrinsic double power-law LF is transformed by the effects of
relativistic beaming. It can be shown that, for astrophysically
relevant values of $a_1$, $a_2$ and $\gamma$ the most general
situation is that in which the beamed LF is a triple broken
power-law \cite{heinz:07}. The faint- and bright-end slopes are unchanged, while
around the knee an intermediate-slope $-(p+1)/p=-3/2$ is found.

The problem of inverting the Urry and Padovani (1991) solution, i.e. of
deriving the intrinsic shape of the radio loud AGN LF given the
observed one and a few assumption about $p$ and $\gamma$ is greatly
simplified if the intermediate part of the beamed 
luminosity function is neglected. In
practice, in such a case the intrinsic, de-beamed, FSLF
 $\phi$ can be obtained from the observed
one simply by the following transformations of the LF knee, while
keeping both end slopes fixed. In detail, following the procedure
described in Urry and Padovani (1991), the LF normalization will be corrected
by a factor
\begin{equation}
\rho' \longrightarrow \rho(z)=\rho'(z) 
\left(\frac{\beta \gamma p C1}{\kappa_{1}}\right) \left(\frac{\kappa
    C1}{C2}\right)^{(a_1-1/(a_2-a_1))};
\end{equation}
while the break luminosity transforms as:
\begin{equation}
L_{\rm R,c}'(z)\longrightarrow L_{\rm R,c}(z)=L_{\rm R,c}'(z) (\kappa C1/C2)^{-1/(a_2-a_1)},
\end{equation}
where $\kappa_{1}=\delta_{\rm max}^{p C1}-\delta_{\rm min}^{p C1}$, 
      $\kappa_{2}=\delta_{\rm max}^{p C2}-\delta_{\rm min}^{p C2}$,
      $\kappa=\kappa_{2}/\kappa_{1}$ and $C1=a_1-(1/p)-1$, 
      $C2=a_2-(1/p)-1$.

In this way, for any given luminosity function and mean Lorentz
factor, it is possible to derive the intrinsic kinetic luminosity function of
the AGN jet, by simply applying eq.~(\ref{eq:kineticlumi}) to the
de-beamed radio luminosity function. 

As opposed to the case of X-ray selected luminosity
functions, the available constraints on the radio cores luminosity
functions are not very stringent. Here we will use two different
determinations, translated into our Cosmology, and computed at
5GHz: the flat spectrum radio
luminosity function from Dunlop \& Peacock (1990) (their HIGH-z
parametrization of their table C4), and that from De
Zotti et al. (2005), given by the sum of flat spectrum radio quasar
and BL Lac objects as in eqs.~(1)-(3) and table 1 of their paper. 
As we will show in the following, and as already discussed in Heinz,
Merloni and Schwab (2007), the total kinetic power released by AGN is
sensitive to the faint-end slope of the kinetic luminosity
function. For consistency with observations of local radio luminosity
functions down to very low luminosities \cite{nagar:05}, 
 we have fixed the $z=0$ faint end slope to $a_1=1.85$
for both the FSLF of Dunlop \& Peacock (1990) and the BL Lac
population of De Zotti et al. (2005). Because $a_1-1$ is (marginally)
 larger than 0.81, the slope of the $W_{\rm kin}$-$L_{\rm R}$ relation
(\ref{eq:radiopower}), the total kinetic power is determined by the
lower cut-off of our radio luminosity function, chosen here to
correspond to the radio luminosity above which the total number of
sources is equal to the total number of SMBH with mass above $5 \times
10^5 M_{\odot}$ (see \S~\ref{sec:doall}). However, the difference between
the total kinetic power obtained in this way and that one would get
for $a_1<1.81$ is smaller than the uncertainty introduced by the
intrinsic scatter in the relation~(\ref{eq:radiopower}). Therefore,
 for the ease of computation, we have allowed a very mild evolution
 (flattening)  
of such a faint-end slope between $z=0$ and $z=2$, where we have
fixed $a_1=1.75$. Indeed, there are independent lines of evidence
pointing towards a flattening of the faint-end of the AGN radio
luminosity function at high redshift \cite{cirasuolo:05}, 
similarly to the well known evolution observed in the X-ray band.
Future better constraints on the high redshift evolution of the faint
radio AGN population will surely help tightening our constraints on
the overall kinetic power density.

The observed luminosity functions are then de-beamed following the
above described prescription, assuming $p=2$. 
For the mean jet Lorentz factor, we adopt
the most likely value based on the statistical analysis of core radio
luminosity of a small
sample of 15 AGN with measured kinetic power (see Merloni and Heinz
2007 for details)\footnote{We note here that the majority
  of studies of jet core velocity structures at VLBI scales have
  recently shown that the distribution of intrinsic Lorentz factors is
relatively broad, typically in the range $3<\gamma<30$,
\cite{laing:99,cohen:07}. Although cumbersome, it is straightforward
to compute the de-beamed radio luminosity function assuming more
complicated distributions of the bulk Lorentz factors. However, for
the sake of clarity in the current discussion, we leave this exercise
to further studies.}: $\gamma=7$. Also, for the sake of simplicity,
and lacking any clear observational evidence, we
do not account here for the possibility of varying Lorentz factors
with redshift.

So far, we have proceeded under the assumption that the observed FSLF
are complete, in the sense that they account for the totality of jet
cores. In fact,
 we have to discuss also the possible contribution from flat
spectrum cores which are 'hidden' below a more powerful, extended
steep--spectrum component. By definition, the
FSLF contains all black holes except those included in the
steep--spectrum luminosity function (SSLF), which
are dominated by optically thin synchrotron emission.  The scaling
relation from eq.~(\ref{eq:radiopower}) does not hold for
extended, steep--spectrum sources (but see B{\^i}rzan et al. 2004;
Best et al. 2006). We can, however, derive an upper limit to
the contribution from steep spectrum sources.  Assuming a typical
optically thin synchrotron spectral index of 0.65, any underlying flat
spectrum component would have to fall below $\sim$20\% of the observed 5GHz
luminosity, otherwise the source would become too flat to qualify as a
steep--spectrum source.  Using the steep--spectrum luminosity
functions (SSLF) from Dunlop \& Peacock (1990; high-z parametrization
of their table C4) and De Zotti et al. (2005), we
then calculate a maximal core luminosity function comprising both flat
spectrum sources and those 'hidden' cores for which we take a
contribution equal to 10\% of the SSLF. This can be considered as an
extreme case in which the objects with powerful (optically thin) radio
lobes have a core-to-lobe flux ratio (the so-called core prominence,
Bridle et al. 1994) of 0.1 (for a discussion on the observational
constraints on the intrinsic core prominence, see Laing et al. 1999,
Jackson and Wall 1999). 

The kinetic luminosity functions calculated from the observed radio
cores luminosity function with the help of eq.~(\ref{eq:radiopower})
are shown in Figure~\ref{fig:lkinf}, where we plot the luminosity
function $\Phi_W$ multiplied by the kinetic power $W_{\rm kin}$ in a
representation that highlights the luminosity range where most of the
power is released. For the sake of clarity we have
chosen to keep separate the results obtained from adopting only the
FSLF (left panel [FS]) from those obtained adding to the FSLF a
contribution of 10\% of the observed SSLF (right plot [FS+SS$_{\rm
  max}$]). As discussed above, we regard this latter as an effective
upper limit to the AGN kinetic luminosity function.
Figure~\ref{fig:lkinf} also shows the decomposition of the total
kinetic luminosity functions into the two kinetically dominated modes
of accretion, as discussed in \S~\ref{sec:modes}. For both cases (FS
and FS+SS$_{\rm max}$), at $z\la 1$, the
kinetic luminosity functions is dominated by low luminosity (LK)
objects, with bright, radio loud QSOs (HK) only contributing
to the high power tail. At, and above, the peak of the BHAR density evolution
($z\sim 1.5-2.5$, see Fig.~\ref{fig:zev_mdot_datasfr}) HK objects start
dominating the total kinetic power output of the growing black holes
population, even if the relative contribution of LK sources is always
substantial. 

\begin{figure*}
\centering
\begin{tabular}{ll}
\psfig{figure=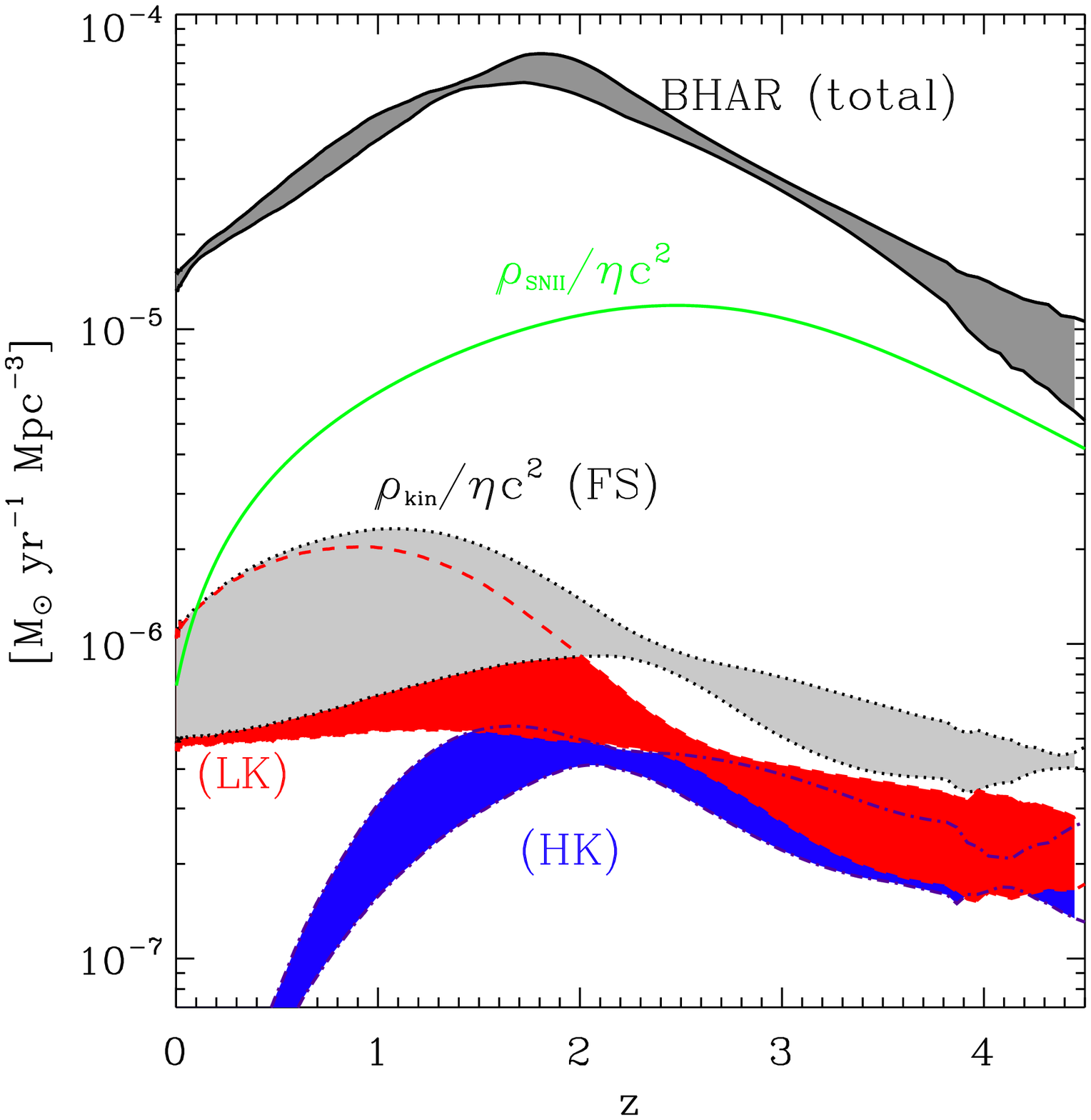,height=8.5cm}&
\psfig{figure=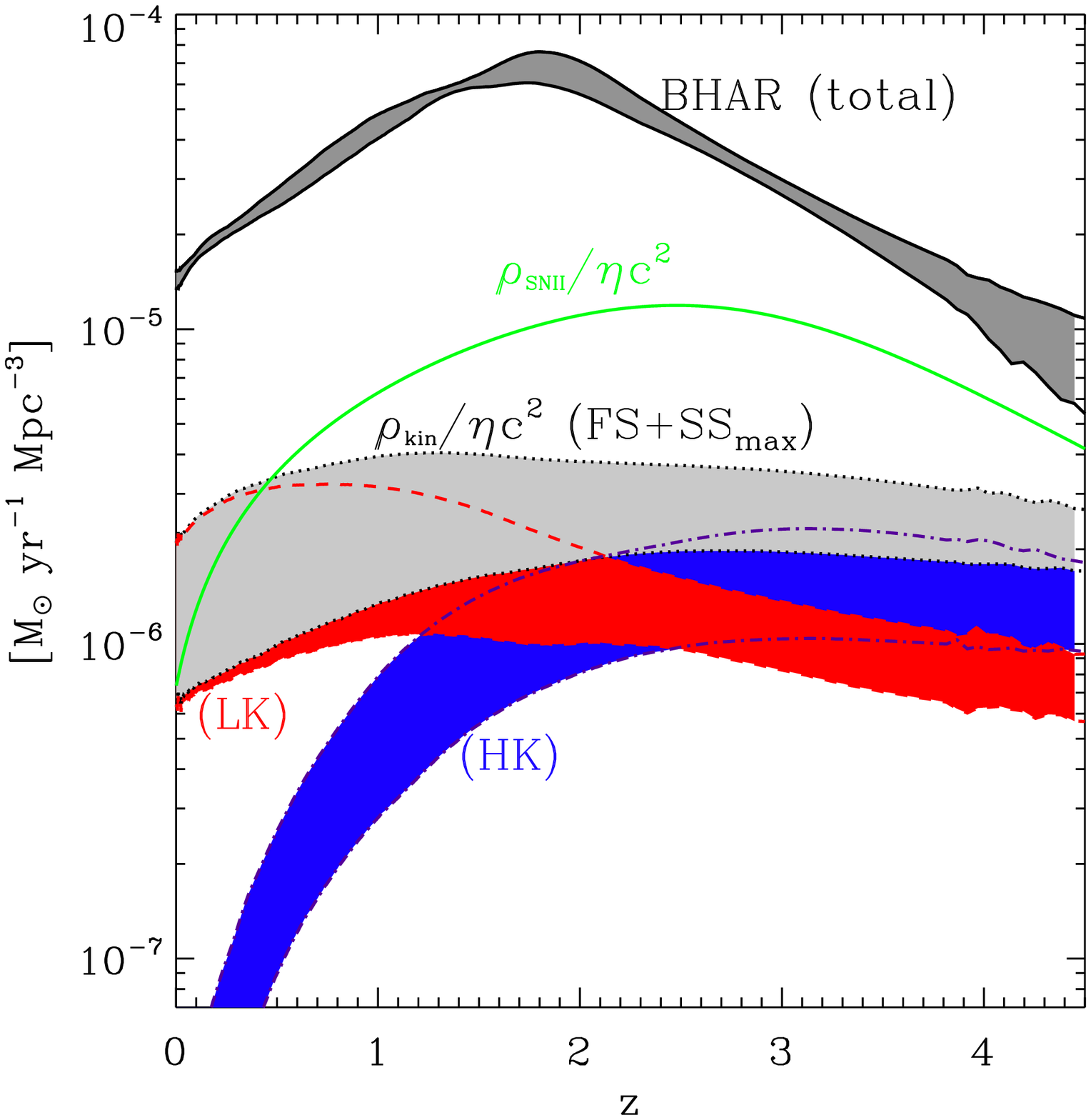,height=8.5cm}\\
\end{tabular}
\caption{Redshift evolution of the kinetic energy density (divided by
  $\eta c^2$ and plotted in units of $M_{\odot}$ yr$^{-1}$ Mpc$^{-3}$)
 is shown with dotted lines and gray area. The top solid lines show
 the BHAR density evolution, $\Psi_{\rm BHAR}(z)$ (see
 fig.~\ref{fig:zev_mdot_083}), while red (dashed lines) and blue
 (dot-dashed lines) bands represent the relative contribution to the
 total kinetic power from LK and HK sources, respectively. The green
 solid line shows the estimated total kinetic power output from SNe
 II, as inferred by Hopkins and Beacom (2006) based on the measured
 star formation rate density. The left plot shows the case in which
 only flat spectrum radio AGN are include in the intrinsic core radio
 luminosity function (FS), while the right hand plot shows the case in
which a maximal contribution from 'hidden' flat spectrum cores in
steep--spectrum sources is included.}   
\label{fig:zevkin_all_083}
\end{figure*}

\subsection{Kinetic energy density evolution and kinetic efficiency}
\label{sec:feedback}
Integrating the kinetic luminosity functions obtained following the
procedure described in the previous section, we arrive to an estimate
of the total kinetic energy density as a function of redshift.

Figure~\ref{fig:zevkin_all_083} shows the redshift evolution of the
total integrated kinetic energy density (divided by $\eta c^2$, so
that it can be displayed in units of $M_{\odot}$ yr$^{-1}$
Mpc$^{-3}$). Once more, we show separately the results of the calculation
done including flat spectrum sources only (left panel) and one in
which a maximal contribution from hidden cores in steep spectrum
sources is added (right panel). The total jet kinetic power density
(shown as grey shaded areas) is split into the contributions from LK
and HK modes of growth. Because the average Eddington ratio increases
towards high redshift, so it does the relative contribution of HK
objects (i.e. radio loud QSOs, or powerful FR II), irrespective of the
choice of core radio luminosity function. As a term of comparison, the
plot shows also the evolution of the total black hole accretion rate
density (BHAR), which we discussed at length in
section~\ref{sec:integral}. Also shown is the estimated total kinetic
power output of SNe II, as computed directly from the best fit to the 
star formation rate density evolution \cite{hopkinsb:06}. The total
kinetic power released by low luminosity (LK) AGN at $z\la 0.3$, where
it is by far the dominant contributor to the total jet kinetic power,
is comparable to that released by type II supernovae, which, however,
become about a factor of few (up to about an order of magnitude)
higher at the peak of star formation activity. 

Given the stark differences in the redshift evolution of low- and
high--power sources, and the fact that the relative importance of
kinetic power output increases as the Eddington ratio decreases (see
\S~\ref{sec:modes}), 
we can conclude that the most important feature of the evolution of
the kinetic energy output of growing black holes, as compared to the
radiative power output (see Figure~\ref{fig:zevrad_083}) is the very
mild redshift evolution. In this respect, it is interesting to mention
that Cavaliere \& Lapi (2008) have
recently shown how the problem of the 'missing baryons' in groups and
cluster of galaxies can be understood in terms of a 'dual' model for
AGN feedback (kinetic at low $z$ and radiative at high $z$) that is in
very good agreement with the overall picture of SMBH growth and energy
output that emerges from the work presented here.

\begin{figure*}
\centering
\begin{tabular}{ll}
\psfig{figure=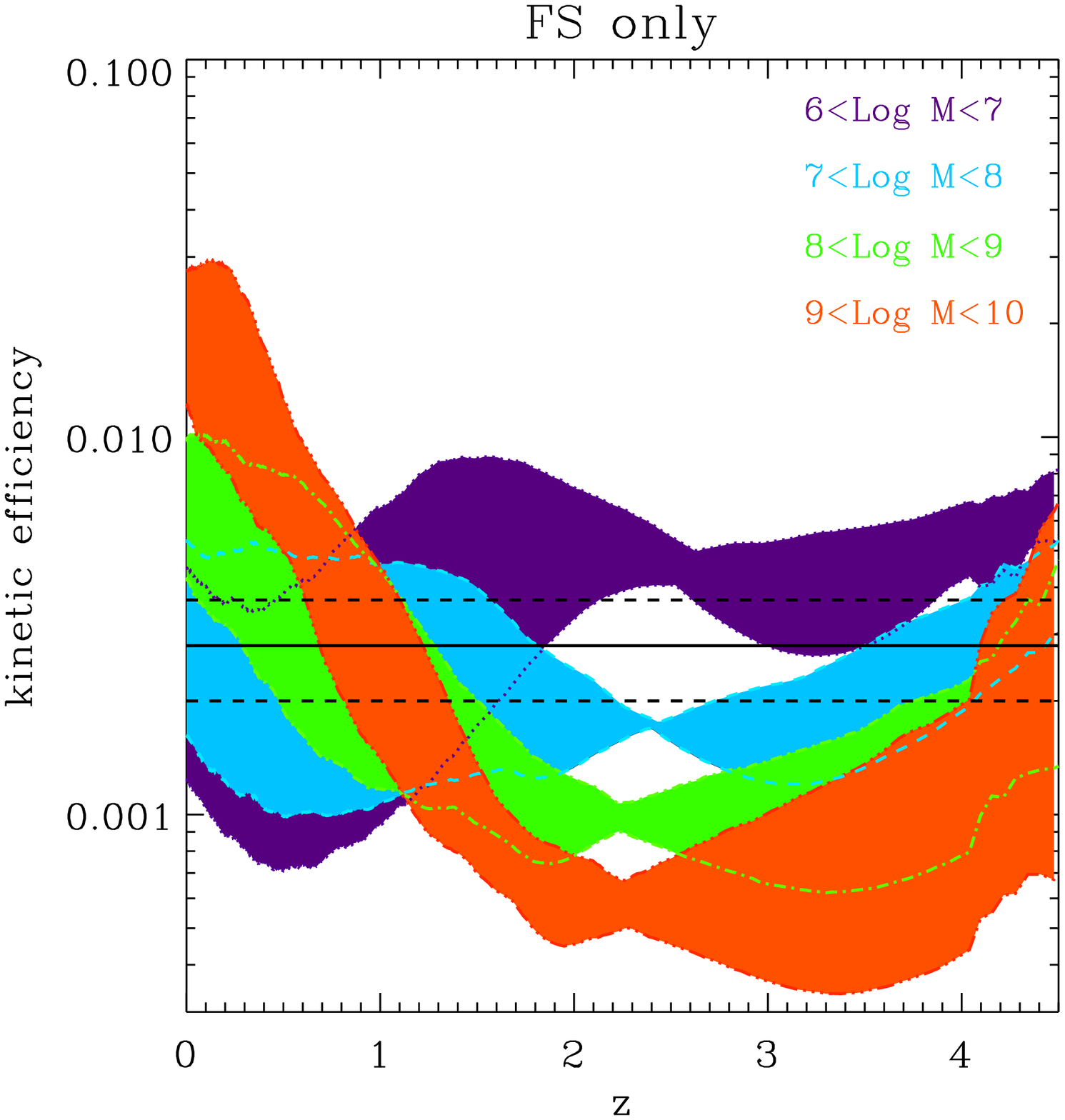,height=8.5cm}&
\psfig{figure=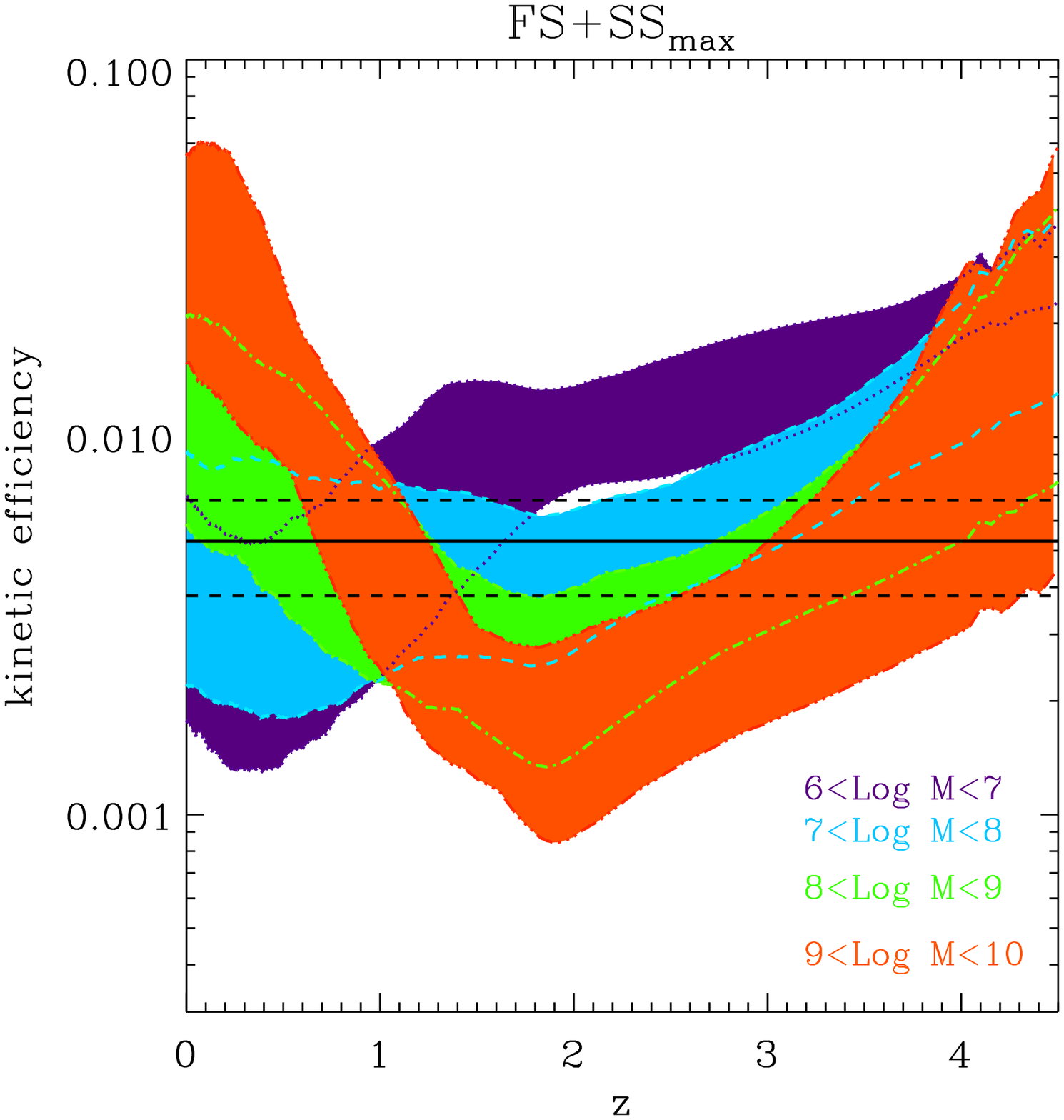,height=8.5cm}\\
\end{tabular}
\caption{Redshift evolution of the kinetic efficiency $\epsilon_{\rm
    kin}$. SMBH of different masses are here plotted separately, with
  a color coding analogous to that of Fig.~\ref{fig:zevrad_083}.
The right hand plot shows the results obtained including flat spectrum
radio sources only, while 
the right hand one shows the results obtained including also a maximal
contribution from the hidden cores of steep spectrum radio AGN. In
each plot, horizontal black solid lines mark the mass weighted average
values for the kinetic efficiencies, with the dashed lines
representing the uncertainties from the particular choice of radio and
X-ray luminosity functions.}
\label{fig:zevratio_083_masses}
\end{figure*}


A fundamental question we are seeking the answer for is the accurate
determination of the kinetic energy production efficiency of
 growing black holes. This bears obvious important consequences for AGN feedback
 models, but has so far been treated as a free parameter. Our
 calculations allow for a more precise determination of such parameter.
Let us first compute the total integrated (mass
 weighted) average kinetic efficiency as:
\begin{eqnarray}
\label{eq:eps_kin}
\langle\epsilon_{\rm kin}\rangle \equiv && \frac {\int_0^{z_i} \rho_{\rm
    kin}(z) dz}{\int_0^{z_i}c^2 \Psi_{\rm BHAR}(z) dz} \nonumber \\ 
&& \simeq \left\{ 
\begin{array}{ll}
0.0028 \pm 0.0008 &   {\rm FS}\\
0.0053 \pm 0.0015 &   {\rm FS+SS_{\rm max}} \\
\end{array} \right. 
\end{eqnarray} 
These numbers are consistent with those found in Heinz, Merloni and
Schwab (2007), once we consider the mean Lorentz factor of $\gamma=7$
used to calculate the intrinsic FSLF, and the steeper slope of the
$L_{\rm R}$--$W_{\rm kin}$ correlation (\ref{eq:radiopower}). On the
other hand, they are somewhat smaller (less than a factor of 2) than
those found in both K{\"o}rding, Jester \& Fender (2008) and in 
Shankar et al. (2008a), where, however, only
steep--spectrum sources were used, and the total kinetic power was derived 
from the correlation between
extended radio luminosity and kinetic power of Willott et al. (1999). 

Combining the results discussed in section~\ref{sec:integral} on the
average radiative efficiency with those shown in
eq.~(\ref{eq:eps_kin}) we conclude that SMBH during their growth from
$z\sim 5$ till now convert about 15-25 times more rest mass energy into
radiation than into kinetic power, with the exact number depending on
the poorly known details of the intrinsic jet cores luminosity
function, as well as on our assumptions about the beaming corrections
to be made. We note here that numerical simulations of AGN feedback in
merging galaxies \cite{dimatteo:05} fix the feedback efficiency to
about $5\times 10^{-3}$ in order to reproduce the $M-\sigma$
relation. In light of what we have shown, it could be argued that this
feedback might be almost entirely provided by kinetically dominated
modes of SMBH growth, as long as the coupling efficiency between the
kinetic power provided by AGN and the interstellar and intra-cluster
medium is equal to one.
 
We can also compute directly the kinetic efficiency as a
function of redshift and SMBH mass, which we show in
Fig.~\ref{fig:zevratio_083_masses}. In each of the two panels,
corresponding to the FS and FS+SS$_{\rm max}$ cases, the horizontal
solid lines show the mass weighted average from
eq.~(\ref{eq:eps_kin}). 
The various curves describe the main
properties of kinetic feedback as we observe it. 
For each of the chosen mass ranges,
the kinetic efficiency has a minimum when black holes of that mass
experience their fastest growth: 
this is a different way of restating the conclusion that most
of the growth of a SMBH happens during radiatively efficient phases of
accretion. However, when the mass increases, SMBH are more and more
likely to enter the LK mode (see section~\ref{sec:modes}), which
increases their kinetic efficiency. More massive holes enter this
phase earlier, and by $z=0$ they have reached the highest kinetic
efficiency of $\sim 2\div5 \times 10^{-2}$, as it is clearly seen in
Figure~\ref{fig:zevratio_083_masses}. This is a natural consequence of
the observed anti-hierarchical growth of the SMBH population, and of
the chosen physical model for the accretion mode of low-Eddington
ratio objects.

\section{Discussion and conclusions}
\label{sec:discussion}
We have presented a comprehensive review of our current knowledge of
the evolution of AGN and of the associated growth of the supermassive
black holes population. 

Similar to the case of X-ray background synthesis models, 
where accurate determinations of the XRB intensity and spectral shape, 
coupled with the resolution of this radiation into
individual sources, allow very sensitive tests of how the AGN luminosity
and obscuration evolve with redshift, 
we have argued that accurate determinations
of the local SMBH mass density and of the AGN (bolometric) luminosity functions,
coupled with accretion models that
specify how the observed AGN radiation translates into a black hole
growth rate, allow sensitive tests of how the SMBH population (its mass
function) evolves with redshift. By analogy, we have named
this exercises `AGN synthesis modelling'. In performing it, we have
taken advantage of the fact that the cosmological evolution of SMBH is
markedly simpler than that of their host galaxies, as individual black
hole masses can only grow with time, and SMBH do not transform into
something else as they grow. Moreover, by identifying active AGN
phases with phases of black holes growth, we can follow the 
evolution of the population by solving a simple
continuity equation~(\ref{eq:continuity}), where the mass
function of SMBH at any given time can be used to predict that at any
other time, provided the distribution of accretion rates as a function
of black hole mass is known (see section~\ref{sec:methods}).

Such a straightforward approach is fully justified as long as SMBH
coalescences, which are likely to occur in the nuclei of merged
galaxies, do not alter significantly the shape of the mass functions.
It is well established, as we have also shown here, that the local
black hole mass density is fully consistent with the mass accumulated
onto SMBH in AGN episodes for reasonable values of the accretion
efficiency (see \S~\ref{sec:integral}). Mergers only redistribute mass
so they do not affect any integral constrain, such as the Soltan
argument, and variations thereof.
Until now, evaluating the relative importance of SMBH mergers in the
mass function evolution has proved itself very difficult \cite{yu:02},
due to the large uncertainties in both merging rates of galaxies
(either theoretically or observationally, see
e.g.  Bundy, Treu \& Ellis 2007; Guo and White 2008; 
Hopkins et al. 2008 and references therein) 
and in the physical processes involved in the merger of
the two nuclear black holes (see Colpi et al. 2008 for a recent review).
Shankar et al. (2008b) have attempted to estimate such effect
considering only equal mass merger, at a rate most likely far in
excess of the true one, and concluded that the effect of SMBH mergers
on the local mass function
is overall smaller than the current uncertainties in the AGN bolometric
luminosity function itself, and may be relevant, if at all, only at the very
high mass end of the distribution. A more detailed comparison, taking
into account the full mass and redshift dependencies of the merger
rates, is however necessary before a firm conclusion can be drawn.

In order to carry out our calculation, we assumed that black
holes accrete in just three distinct physical states, or ``modes'':
a radiatively inefficient, 
kinetically dominated mode at low Eddington ratios (LK, the so-called
``radio mode'' of the recent literature), and two modes at 
high Eddington ratios: a purely radiative one (radio quiet,
HR), and a kinetic (radio loud, HK) mode, with the former outnumbering
the latter by about a factor of 10. Such a classification is based on
our current knowledge of state transitions in stellar mass black hole
X-ray binaries as well as on a substantial body of works on scaling
relations in nearby SMBH (see \S~\ref{sec:modes}). It allows a
relatively simple mapping of the observed luminosities (radio cores,
X-ray and/or bolometric) into the physical quantities related to any
growing black hole: its accretion rate and the released kinetic power.

With the aid of such a classification, we have solved the
continuity equation for the black hole mass function using the locally
determined one as a boundary condition, and the 
hard X-ray luminosity function as tracer of AGN growth rate
distribution, supplemented with luminosity-dependent bolometric
corrections \cite{marconi:04} 
and absorbing column density distributions. 
By construction, our
adopted intrinsic hard X-ray luminosity functions satisfy the XRB constrain (as
well as the sources number counts, and many others), as
they adopt the most recent synthesis model as a input (see Gilli et
al. 2007 for details). One remaining uncertainty with these models
lies in the actual luminosity and redshift distribution of heavily
obscured (Compton thick) AGN. However, according to the adopted column density
distribution, the contribution to the overall mass growth from Compton
thick AGN amounts to just about 20\%. Firmer conclusions on this issue will
need more accurate determination of number density and typical
luminosity of these heavily absorbed (and thus invisible in the 2-10
keV band) sources at redshift between 1 and 2. In fact, preliminary results
based on mid-infrared selection criteria \cite{daddi:07,fiore:08} 
seem to show quite a good
agreement with the predictions of the XRB synthesis models adopted
here. 

The main results of our study are summarized below.

\begin{figure}
\centering
\psfig{figure=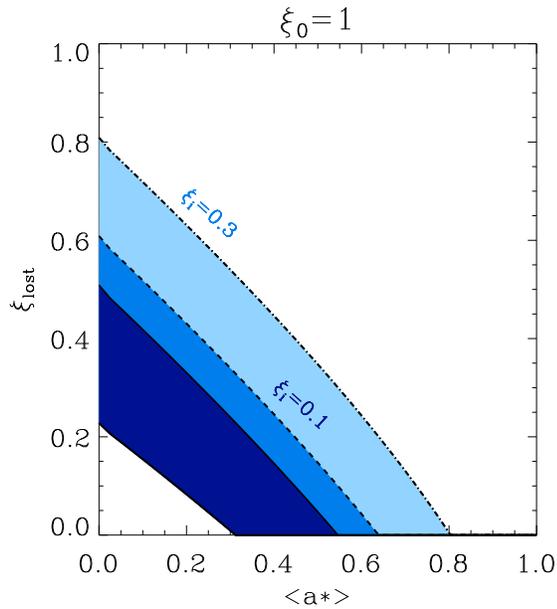,height=8.5cm}
\caption{Joined constraints on the fractional mass density of
  ``wandering'' SMBH, $\xi_{\rm lost}$, ans on the mass weighted spin
  of SMBH, $\langle a*\rangle$, calculated for $\xi_0=1$. The dark
  blue area between solid lines is the allowed area for $\xi_i=0$,
  while the areas between the lower solid line and the upper dashed
  (dot-dashed) line represent the allowed area for $\xi_i=0.1$ ($\xi_i=0.3$).}
\label{fig:xi_a}
\end{figure}

\begin{itemize}

\item{We have shown how using
standard Soltan (1982) type of arguments, i.e. comparing the local
mass density to the integrated mass growth in AGN phases, very tight
constraints can be put on the average radiative efficiency of the
accretion process: $\frac{0.065}{\xi_0(1+\xi_{\rm lost})}\la
\epsilon_{\rm rad} 
 \la \frac{0.070}{\xi_0(1-\xi_i+\xi_{\rm lost})}$, where $\xi_0$ is the local
mass density in units of 4.3 $\times 10^{5} M_{\odot}$ Mpc$^{-3}$,
while $\xi_i$ and $\xi_{\rm lost}$ are the mass density of $z\sim 5$ and
``wandering'' SMBH, respectively (also in units of the local mass
density).
Figure~\ref{fig:xi_a} shows the constraints derived from
eq.~(\ref{eq:radeff_const}) in the $\xi_{\rm lost}$, $\langle
a*\rangle$ panel, where $\langle a*\rangle$ is the mass weighted
average spin parameter of the SMBH calculated inverting the classical
GR $\eta(a)$ relation (Shapiro 2005) and taking into account the
assumed relationship of eq.~(\ref{eq:radeff}) between $\eta$ and
$\epsilon_{\rm rad}$.
It is interesting to note that both the amount (numbers and masses) of
black holes effectively ejected from galactic nuclei due to
gravitational wave recoil after a merger and the average radiative
efficiency of accreting black holes depend on the spin distribution of
evolving SMBH (see e.g. Volonteri 2007, Berti and Volonteri 2008). 
Thus, eq.~(\ref{eq:radeff_const}) couples implicitly the spin
distribution of accreting black holes (through $\xi_{\rm lost}$ and
$\epsilon_{\rm rad}$) and the properties of the seed
black hole population, whose density must be reflected in the $z\sim
5$ mass density $\xi_i$. At face value, our results indicate that, for
$\xi_0=1\gg \xi_i$, SMBH must have on average a rather low spin,
similar to what predicted by models of AGN fuelling that take into
account the self-gravity of an accretion disc and result in accretion
events which are small and randomly-oriented, thus favoring effective
spin-down of the hole \cite{king:08}. }

\item{We confirm previous
results and clearly demonstrate that, at least for $z\la 1.5$, SMBH mass
function evolves anti-hierarchically, i.e. the most massive holes grew
earlier and faster than less massive ones (\S~\ref{sec:mdotf}). 
By looking at the distribution of SMBH growth times as a function of
mass and redshift, for the first time we find
hints of a reversal of such a downsizing behaviour at redshifts close
to and above the peak of 
the black hole accretion rate density ($z\approx 2$), where we witness
the epoch of rapid growth of almost the entire SMBH population. 
Our results bring the study of AGN 'downsizing' from a
phenomenological level, in which one simply describes the observed
behaviour of the luminosity functions \cite{hasinger:05,bongiorno:07},
to a more physical one, in which such an observed behaviour is
explained in terms of evolution of physical quantities (mass and
accretion rate), which could be more directly related to the parallel
evolution of the host galaxies population.}

\item{Differently from most
previous analytic, semi-analytic and numerical models of black holes growth, we
do not assume any specific distribution of Eddington ratios, rather we
determine it empirically by coupling the mass and luminosity
functions and the set of fundamental relations between observables in
the three accretion modes. SMBH always show a very broad accretion rate
distribution, and we have highlighted the profound consequences of this fact
for our understanding of observed AGN fractions in galaxies, as well as
on the empirical determination of SMBH mass functions from large
surveys (\S~\ref{sec:active}).} 

\item{We have presented the
cosmological evolution of the kinetic luminosity function of AGN, based
on the evolution of their flat spectrum radio luminosity function and
on the empirical correlation between intrinsic radio core luminosity
and kinetic power found in Merloni and Heinz (2007). As opposed to the
mass growth evolution, the kinetic luminosity function so derived is
not very tightly constrained due to poor observational information on the true
(intrinsic) radio core luminosity function at high redshift and to the
uncertain distribution of the AGN jet bulk Lorentz factors needed to
correct for relativistic beaming of their core emission. Nevertheless,
we were able to measure the overall
efficiency of SMBH in converting accreted rest mass energy into
kinetic power, the ``kinetic efficiency'' $\epsilon_{\rm kin}$,
 which ranges between 3 and 5 $\times 10^{-3}$, depending on
the choice of the radio core luminosity function. This value is
 somewhat smaller (but less than a factor of 2) than
those found independently in both K{\"o}rding, Jester \& Fender (2008) and in 
Shankar et al. (2008a), where only
steep--spectrum sources were used, and the total kinetic power was derived 
from the correlation between
extended radio luminosity and kinetic power of Willott et
al. (1999). Nevertheless, 
the redshift distribution of the kinetic power density released by
growing SMBH is in good agreement with that derived by K{\"o}rding et
al. (2008), lending further support to the idea that current measures
of the kinetic AGN power are reaching quite robust conclusions.} 

\item{ As for the local ($z=0$) AGN kinetic power density, $\rho_{\rm kin}$, 
we found it ranges between 
2.5 and 10 $\times$ 10$^{39}$ erg s$^{-1}$ Mpc$^{-3}$, comparable with
the total kinetic power density from type II Supernovae ($\rho_{\rm
  SNII}\simeq 4 \times 10^{39}$  erg s$^{-1}$ Mpc$^{-3}$, Hopkins and
Beacom 2006) while the local
stellar luminosity density is about  $\rho_{*}\simeq 2 \times 10^{41}$
erg s$^{-1}$ Mpc$^{-3}$ and the total AGN radiative density is about
$\rho_{\rm rad}\simeq 1.6 \times 10^{40}$ erg s$^{-1}$ Mpc$^{-3}$.}

\item{Our capability to resolve the mass and accretion rate functions 
allows us to separate the evolution of both growth rate and kinetic
energy density into different mass bins and into the various modes of
accretion. In doing so, we found that most (73-80\%) of the local mass
density was accumulated during radiatively efficient modes (either HR
or HK), with the remaining 20-27\% in the radiatively inefficient LK
mode. The kinetic power density at low redshift is
completely dominated by low luminosity AGN, while the contribution
from radio loud QSOs (mode HK) becomes significant at $z\sim2$
(see section~\ref{sec:feedback}).  The measured $\epsilon_{\rm kin}$
varies strongly with SMBH mass and redshift, being maximal for very
massive holes at late times, a property in agreement with what
required for the AGN feedback by many recent galaxy formation models.}

\end{itemize}

The richness of details we have been able to unveil for the
cosmological evolution of supermassive black holes demonstrates that 
times are ripe for comprehensive unified approaches to
the multi-wavelength AGN phenomenology. At the same time, our results
should serve as a stimulus for semi-analytic and numerical modelers of
structure formation in the Universe to consider more detailed
physical models for the evolution of the black hole population.

\section*{Acknowledgments}
We thank E. Berti, S. Bonoli, M. Brusa, G. De Lucia, P. Hopkins, 
G. Hasinger, K. Iwasawa, S. Kochfar, J. Malzac and
J. Silverman for useful discussions and suggestions.
SH acknowledge support through NASA grant GO7-8102X

\label{lastpage}

\end{document}